\documentclass[a4paper,11pt]{article}

\usepackage{jheppub}

\usepackage[utf8]{inputenc}
\usepackage{amsmath}
\usepackage{graphicx}
\usepackage{color}
\usepackage[dvipsnames]{xcolor}
\usepackage[version=3]{mhchem}

\definecolor{nicegreen}{rgb}{0.1,0.5,0.1}
\definecolor{darkblue}{rgb}{0.15, 0.2, .85}
\definecolor{darkgreen}{rgb}{0.1,0,0.3}
\definecolor{darkred}{rgb}{0.6,0,0}
 
\newcommand{\sinsq}[1]{\sin^2 \theta_{#1}} 
\newcommand{\Deltamsq}[1]{\Delta m^2_{#1}}
\newcommand{\eV}{\, \rm eV}
\newcommand{\nue}{\nu_e}
\newcommand{\nuebar}{\overline{\nu}_e}
\newcommand{\numu}{\nu_\mu}
\newcommand{\numubar}{\overline{\nu}_\mu}
\newcommand{\rsun}{r_\odot}
\newcommand{\RMW}{R_{\rm MW}}
\newcommand{\smax}{s_{\rm max}}
\newcommand{\mdm}{m_{\rm DM}}
\newcommand{\Ekin}{E_{\rm kin}}
\newcommand{\Evis}{E_{\rm vis}}
\newcommand{\Javg}{J_{\Delta\Omega}}
\newcommand{\kpc}{\, {\rm kpc}}
\newcommand{\MeV}{\, {\rm MeV}}
\newcommand{\GeV}{\, {\rm GeV}}

\newcommand{\cm}{\, {\rm cm}}
\newcommand{\s}{\, {\rm s}}

\newcommand{\zGC}{z_{\rm GC}}

\newcommand{\sigmav}{\langle \sigma v \rangle}
\newcommand{\GENIE}{{\tt GENIE }}

\definecolor{ShamrockGreen}{rgb}{0.0, 0.62, 0.38}

\title{Searching for Sub-GeV Dark Matter in the Galactic Centre using Hyper-Kamiokande}

\author[a]{Nicole F. Bell,}
\author[a]{Matthew J. Dolan,}
\author[a]{Sandra Robles}
\affiliation[a]{ARC Centre of Excellence for Dark Matter Particle Physics, \\
School of Physics, The University of Melbourne, Victoria 3010, Australia}
\emailAdd{n.bell@unimelb.edu.au}
\emailAdd{matthew.dolan@unimelb.edu.au}
\emailAdd{sandra.robles@unimelb.edu.au}

\abstract{
Indirect detection of dark matter via its annihilation products is a key technique in the search for dark matter in the form of weakly interacting massive particles (WIMPs). Strong constraints exist on the annihilation of WIMPs to {\it highly visible} Standard Model final states such as photons or charged particles.  In the case of s-wave annihilation, this typically eliminates thermal relic cross sections for dark matter of mass below $\cal{O}$(10)~GeV.  However, such limits typically neglect the possibility that dark matter may annihilate to {\it assumed invisible} or hard-to-detect final states, such as neutrinos.
This is a difficult paradigm to probe due to the weak neutrino interaction cross section. Considering dark matter annihilation in the Galactic halo, we study the prospects for indirect detection using the Hyper-Kamiokande (HyperK) neutrino experiment, for dark matter of mass below 1 GeV. We undertake a dedicated simulation of the HyperK detector, which we benchmark against results from the similar Super-Kamiokande experiment and HyperK physics projections. We provide projections for the annihilation cross-sections that can be probed by HyperK for annihilation to muon or neutrino final states, and discuss uncertainties associated with the dark matter halo profile. For neutrino final states, we find that HyperK is sensitive to thermal annihilation cross-sections for dark matter with mass around 20~MeV, assuming an NFW halo profile. We also discuss the effects of neutron tagging, and prospects for improving the reach at low mass.}

\begin{document}
\maketitle

\section{Introduction}
\label{sec:intro}

The nature of dark matter (DM) is an enduring mystery. In many theories, dark matter can interact with some of the particles of the Standard Model (SM), which provides both a DM production mechanism in the early Universe and a way of detecting DM in the Universe today. In many of these scenarios, astrophysical dark matter can annihilate to SM particles, raising the possibility indirectly detecting DM through searches for these annihilation products. The natural targets for such searches are regions of high dark matter density such as the Galactic Centre or dwarf spheroidal galaxies, and many limits based on these searches have been published. A recent analysis indicates that combining these results for thermal dark matter, with the assumption that the annihilation is a $2\to 2$ $s$-wave process, leads to a lower bound on the DM mass of $m_\chi \gtrsim 20$~GeV~\cite{Leane:2018kjk}. However, that analysis included only visible final states, such as photons, charged particles or other highly detectable SM particles.

To comprehensively test the WIMP paradigm, one must also consider those annihilation products which may be harder to detect. This includes states which might be assumed as largely invisible, such as neutrinos, together with truly invisible states, such as other dark sector particles. Though the latter cannot be excluded, the requirement that dark matter is thermally produced in the early Universe argues for a coupling to the SM. Although neutrinos are typically the most difficult to detect SM annihilation product and hence lead to conservative limits~\cite{Beacom:2006tt,Yuksel:2007ac}, we show that they are actually detectable at an interesting sensitivity.

For dark matter that annihilates to neutrinos, the lower bound on $m_\chi$ is currently set by measurements of the effective number of neutrinos, $N_{\rm{eff}}$, from the Cosmic Microwave Background and Big Bang Nucleosynthesis. This leads to a lower bound on the mass of neutrinophilic dark matter of between 3.7 and 9.4~MeV, depending on the degrees of freedom of the DM~\cite{Boehm:2013jpa,Nollett:2013pwa,Nollett:2014lwa,Escudero:2018mvt,Sabti:2019mhn}. Future precision CMB experiments such as the Simons Observatory and CMB-S4 may increase these lower bounds to 10-16~MeV. There are well-defined UV-complete models corresponding to this region of DM parameter space, for instance~\cite{Boehm:2013jpa,Campo:2017nwh,Elor:2018twp,Blennow:2019fhy,Ballett:2019pyw}. DM-neutrino interactions may also have relevance for structure formation in the early Universe~\cite{Boehm:2000gq,Mangano:2006mp,Wilkinson:2014ksa} and can be constrained through cosmological measurements.

Neutrino final states are particularly challenging for indirect detection due to their very weak interaction cross-sections. However, limits on dark matter annihilating to neutrinos have been set by neutrino experiments such as Super-Kamiokande (SuperK)~\cite{Frankiewicz:2015zma,Frankiewicz:2017trk} IceCube~\cite{Aartsen:2015xej} and ANTARES~\cite{Albert:2016emp}, covering masses from 1~GeV up to 100~TeV. Other experiments including HESS~\cite{Abdallah:2016ygi} and Fermi~\cite{Ackermann:2015zua} have presented limits based on searches for final states such as $\mu^+\mu^-$ and $W^+ W^-$, which lead to neutrinos via their decays. 

We shall see that the neutrino experiments have sensitivity to relic density scale annihilation cross sections in a region of parameter space that overlaps with that probed by the CMB and BBN measurements. However, these very different approaches provide important complementarity. In particular, the neutrino experiments are direct in the sense that the annihilation products are actually detected. Importantly, in the case that a signal were to be observed, they would be able to provide strong evidence for a DM origin and a determination of the DM mass.

While limits set by these collaborations start at masses of 1~GeV, a number of groups have re-interpreted data and searches for other phenomena to constrain DM-neutrino interactions below this mass. These include measurements from the BOREXINO solar neutrino observatory~\cite{Bellini:2010gn,Campo:2017nwh,Arguelles:2019ouk}, KamLAND~\cite{Collaboration:2011jza,Arguelles:2019ouk} and Super-Kamiokande~\cite{PalomaresRuiz:2007eu,Campo:2017nwh,Klop:2018ltd,Arguelles:2019ouk}. Other earlier work in this area includes~\cite{Beacom:2006tt,Yuksel:2007ac,Rott:2011fh,Kappl:2011kz,Primulando:2017kxf}. An up-to-date and comprehensive summary of the current limits on DM annihilating to neutrinos over a wide range of masses is~\cite{Arguelles:2019ouk} and a broad discussion of BSM opportunities at future neutrino experiments can be found in ref.~\cite{Arguelles:2019xgp}.

Between 10~MeV and 1~GeV the strongest limits on DM annihilation into neutrinos are currently set by recasting a variety of results of the  Super-Kamiokande experiment~\cite{Arguelles:2019ouk}. SuperK is a 50~kT water Cherenkov detector at the Kamioka mine site in Japan which, within the next decade, will be superseded by  a new large water Cherenkov detector, Hyper-Kamiokande. HyperK will have exceptional sensitivity to light DM annihilating into neutrinos. However, the HyperK Design Report (DR)~\cite{Abe:2018uyc} does not provide projections for DM masses below 1~GeV, even though the detector threshold will be a few MeV. The purpose of this paper is to estimate the sensitivity of the HyperK experiment to light dark matter annihilating in the Galactic Centre. We will focus on annihilation into neutrinos and muons. 

To do this we use a simulation of the HyperK detector that we describe in detail below. Exploiting the fact that HyperK is built upon the intellectual and technical foundations of SuperK, we first simulated the SuperK detector, validating our simulation using published SuperK results. We then scaled up our SuperK simulation to the dimensions and performance efficiency of HyperK, comparing when possible with results in the HyperK DR. We note that the upcoming JUNO and DUNE experiments, using liquid scintillator and liquid argon respectively, will also have excellent sensitivity to light DM that annihilates into final states involving neutrinos~\cite{Klop:2018ltd,Arguelles:2019ouk}.

The primary backgrounds for dark matter searches in this mass range are from atmospheric neutrinos. However, for DM masses below 100~MeV there is an added complication due to the presence of the Diffuse Supernova Neutrino Background (DSNB). Measurement of the DSNB is much sought after in order to gain information on the star formation rate in the early Universe, for reviews see~\cite{Beacom:2010kk,Mirizzi:2015eza}. However, its presence has not yet been confirmed~\cite{Bays:2011si,Zhang:2013tua}. It is expected that the addition of small quantities of Gadolinium to the water in SuperK and HyperK (which allows neutrons produced in the inverse-beta process to be tagged~\cite{Beacom:2003nk}) will allow the DSNB to be measured~\cite{Horiuchi:2008jz}. In this work, in order to set a conservative limit, we consider the DSNB to be fixed and contribute to the background. The limits that can be set by HyperK in the mass range 20-80~MeV have previously been estimated in ref.~\cite{Campo:2018dfh} by rescaling and re-interpreting previous SuperK DSNB searches. We find that our projections agree quite well with those results. 

An important unknown that affects the translation of a limit on the flux to a constraint on the annihilation cross section, is the  uncertainty associated with the DM halo profile. To estimate the impact of this uncertainly, we present results for three different profiles: the standard NFW~\cite{Navarro:1995iw}, as well as Moore~\cite{Moore:1999gc} and Isothermal~\cite{Bahcall:1980} profiles.
Rather than focusing on a small signal region located at the Galactic Centre, we shall derive conservative all-sky limits that use the DM annihilation signal from the whole halo.  While the HyperK angular resolution is relatively poor in the low mass (sub-GeV) energy regime that is our focus here, we note that our conservative limits could be strengthened somewhat by considering a smaller angular region~\cite{Yuksel:2007ac}.
 
We begin by presenting details of our detector simulation and validation procedure in Section~\ref{sec:sims}. In Section~\ref{sec:limits} then we discuss our event selection, statistics and limit-setting procedures,  to determine projections for the future HyperK reach. The appendix contains ancillary materials including the effects of varying the DM halo model.

\section{HyperK Simulation and Validation}
\label{sec:sims}

In this section we describe our analysis setup and workflow, including the SuperK and HyperK detector simulations. We first discuss the detector geometries and event generation, our implementation of the tracking and smearing of particle momenta and energies, the relevant neutrino interactions and oscillations and, finally, how we categorise events. We then present results validating our workflow and simulations against results from SuperK and the HyperK Design Report.

\subsection{Detector Geometry and Event Generation}
\label{sec:geometry}

The Super-Kamiokande experiment is a large, cylindrical, water Cherenkov detector located in the Mozumi mine in Japan. Hyper-Kamiokande is a next-generation experiment that will be the successor to SuperK and will be located in the nearby Tochibora mine.

The SuperK detector is divided into an inner detector (ID) and outer detector (OD). The inner detector has a volume of 32 kilotons of water, and is surrounded by the outer detector, an approximately two metre wide cylindrical shell that is used mainly for veto purposes. For event classification purposes the SuperK collaboration also define a fiducial volume, a sub-region of the inner detector. For early analyses, this was the region of the ID more than two metres from the ID wall, although improved reconstruction techniques allowed an increase in the fiducial volume in more recent analyses~\cite{Jiang:2019xwn}. The wall of the inner detector is instrumented with photo-multiplier tubes to capture the Cherenkov light. We will not individually model these in our simulation  but instead parametrise their collective overall response to different kinds of neutrino interactions.

The HyperK detector is planned to be constructed in a similar way to SuperK, but on a larger scale. In particular, the size of the fiducial volume will be nearly an order of magnitude larger at HyperK. Our HyperK detector simulation is based on a scaled-up version of our SuperK detector simulation, both of which have been implemented using the ROOT geometry package~\cite{Brun:1997pa}.  The detector dimensions for SuperK correspond to SuperK-IV~\cite{Fukuda:2002uc,Abe:2013gga} while those for HyperK are taken from the HyperK DR~\cite{Abe:2018uyc}, as detailed in Table~\ref{tab:detectorparameters}. In our simulations, the properties of compound materials such as stainless steel, concrete and air have been taken from GEANT4~\cite{Brun:1994aa}.
\begin{table}[h]
  \centering
  \begin{tabular}{|l|c|c|}
   \hline
                        & SK & HK-1TankHD  \\
   \hline
   Depth   &    1000 m  &    650 m  \\
   Tank diameter &   39 m & 74 m \\
   Tank height &   42 m & 60 m \\
   Total volume          &  50 kton & 258 kton \\
   Fiducial volume        &   22.5 kton & 187 kton \\
   Outer detector thickness  & $\sim$ 2 m & 1-2~m \\
   \hline
  \end{tabular}
  \caption{Dimensions of the SuperK~\cite{Fukuda:2002uc,Abe:2013gga} and future
    HyperK~\cite{Abe:2018uyc} water Cherenkov  detectors. SuperK has undergone several configuration changes during its lifetime. The parameters in the table refer to SuperK-IV. 
  }\label{tab:detectorparameters}
 \end{table}
 
 When neutrinos pass in the vicinity of one of the detectors they can interact with the surrounding rock, the detector material, or the water in the detector. To model these interactions we use the \texttt{GENIE}~3.0.4a~\cite{Andreopoulos:2009rq,Andreopoulos:2015wxa} Monte Carlo package, including the atmospheric neutrino flux and detector geometry drivers\footnote{\texttt{GENIE}~3.0.4 originally had an important bug in the calculation of the CCQE cross-section at low energies. This was remedied in v3.0.6, which appeared while this work was in progress and a patched version, v3.0.4a, which we use.}. We have studied two different tunes, G18\_02a\_00\_000 and G18\_10a\_00\_000, and will refer to these as G18\_02a and G18\_10a in what follows. \texttt{GENIE} includes precomputed cross-sections for neutrino interactions with matter for neutrino energies between 10~MeV and 1~TeV.

 For most masses, atmospheric neutrino are the dominant background for dark matter searches at SuperK and HyperK. Below neutrino energies of 100~MeV there is also an important  contribution from the Diffuse Supernova Neutrino Background (DSNB). While this component has not yet been measured, it is expected that the addition of gadolinium will enable its discovery at SuperK~\cite{Horiuchi:2008jz}. We do not simulate the DSNB in this paper. Instead, we take the HyperK DR projection of the expected DSNB signal for a 10 year experimental running time from Fig.~188 of~\cite{Abe:2018uyc} and subtract the  off the backgrounds. We use remainder as our DSNB background.
 
Another important background below energies of 10~MeV is from solar neutrinos. Since this energy is less than 17~MeV, where the muon spallation background becomes dominant (see discussion below), we do not consider solar neutrinos further.

We use the atmospheric neutrino fluxes calculated by HKKM~\cite{Honda:2011nf}, hereafter HKKM11, presented as a function of azimuth and zenith angle without oscillations at Kamioka. The HKKM11 fluxes are computed only down to energies of 100~MeV. Below that we use the FLUKA~\cite{Battistoni:2005pd} results, which extend down to 13~MeV. The FLUKA results are angle-averaged. To regain some angular information we make the assumption that the angular dependence of the neutrino flux below 100 MeV is the same as at 100 MeV, and distribute the total flux calculated with FLUKA in the same way as the lowest of the HKKM11 energy bins. We obtain the neutrino energy spectra for annihilating dark matter from \texttt{DarkSUSY}~\cite{Gondolo:2004sc,Bringmann:2018lay}. Since we are exclusively interested in sub-GeV dark matter in this paper we neglect possible corrections from electroweak Bremsstrahlung.

We also account for the impact of neutrino oscillations. The neutrino oscillation length is 
\begin{eqnarray}
    L_{\rm osc} &=& \frac{4\pi E}{\delta m^2}  \\
                &=& 2.5 \times 10^6 ~{\rm km}~\frac{(10~{\rm meV})^2}{\delta m^2} \left(\frac{E}{100~{\rm GeV}}\right) \label{eq:osc_solar} \\
                &=& 10  ~{\rm km} ~ \frac{(50~{\rm meV})^2}{\delta m^2} \left(\frac{E}{10~{\rm MeV}}\right). \label{eq:osc_atm}
    \end{eqnarray}
The values in Eq.~\ref{eq:osc_solar} correspond to oscillations of high energy (100 GeV) neutrinos driven by the solar mass splitting, $\delta m_{21}^2 \sim 7 \times 10^{-5} {\rm eV}^2$, while those in Eq.~\ref{eq:osc_atm} correspond to oscillations of lower energy (10 MeV) neutrinos, driven by the atmospheric mass splitting $\delta m_{32}^2 \sim 2.5 \times 10^{-3} {\rm eV}^2$, and encompass the oscillation lengths of relevance for us. Hence, for the neutrino energies that we will consider, the oscillation lengths are of $\mathcal{O}(10-10^6)~\rm{km}$.

This is to be compared with the distance to the Galactic Centre of 8~kpc $\sim 10^{17}$~km.  Even if we were to consider DM annihilation in a very small region near the Galactic Centre, say the inner 1~pc~$\sim 3\times 10^{13}$~km, averaging over the production region should wash out the oscillations. We therefore expect that there will be no oscillatory features in the annihilation flux at Earth. Hence, the final flavour structure of the neutrino flux at Earth can be obtained from that at production using the simple expression
\begin{eqnarray}
    \phi^{\rm final}_{\nu_\alpha}(E) & = \sum\limits_{i,\beta} \phi^{\rm source}_{\nu_\beta} (E) |U_{\beta i}|^2 |U_{\alpha i}|^2,
    \label{eq:DMneuoscvac}
\end{eqnarray}
where $\alpha$ labels flavour states, $i$ labels mass states, and $U_{\alpha i}$ are the PMNS matrix elements,
\begin{equation}
U=\begin{bmatrix}
   c_{13}c_{12} & c_{13}s_{12} & s_{13}e^{-i\delta} \\
   -c_{23}s_{12}-s_{23}s_{13}c_{12}e^{i\delta} & c_{23}c_{12}-s_{23}s_{13}s_{12}e^{i\delta} & s_{23}c_{13} \\
   s_{23}s_{12}-c_{23}s_{13}c_{12}e^{i\delta} & -s_{23}c_{12}-c_{23}s_{13}s_{12}e^{i\delta} & c_{23}c_{13}
  \end{bmatrix}.
\end{equation}
This conclusion will be valid only in the case of downward going neutrinos arriving from the Galactic Centre. For neutrinos that travel upward through the Earth before detection, we should consider matter effects in the Earth. 

We calculate the neutrino oscillation probabilities at the detector depth using the \texttt{nuCraft} code~\cite{Wallraff:2014qka}. The \texttt{nuCraft} code numerically solves the Schr\"odinger equation for neutrinos propagating through the Earth, which is treated as a sphere with smoothly varying mass density, instead of a series of shells of constant matter density. The Earth data come from the Preliminary Reference Earth Model~\cite{Dziewonski:1981xy}. We take the oscillation parameters from the PDG~\cite{Tanabashi:2018oca}, assuming a normal ordering of the neutrino masses. The precise values are shown in Table~\ref{tab:oscparams}.

\begin{table}[h]
\centering
\begin{tabular}{|c|c||c|c|}
\hline
Parameter & Value & Parameter & Value \\
\hline
$\sinsq{12}$ & $0.307 \pm 0.013$ & $\Deltamsq{21}$ & $(7.53 \pm 0.18) \times 10^{-5} \eV^2$  \\
$\sinsq{23}$ & $0.512^{+0.019}_{-0.022}$  & $\Deltamsq{32}$ & $(2.444 \pm 0.034) \times 10^{-3} \eV^2$ \\
$\sinsq{13}$ & $0.0218 \pm 0.0007$ & $\delta$ & $1.37^{+0.18}_{-0.16}$~rad\\
\hline
\end{tabular}
\caption{Neutrino parameters from~\cite{Tanabashi:2018oca} used to oscillate the neutrino fluxes. }
\label{tab:oscparams}
\end{table}

\subsection{Neutrino Interactions}
\label{sec:neutint}

Neutrinos interact with matter through charged-current (CC) interactions in a number of different channels that lead to events within the inner detector. At energies above approximately 10~GeV the total cross-section is dominated by deep inelastic scattering (DIS), where the interaction is directly with the quarks and gluons that constitute the nucleus.

More important for our study are quasi-elastic scattering (QE) and resonance neutrino production. The first of these is the interaction of a neutrino on a bound nucleon leading to $\nu_l + n \to l^- +p$. This is the dominant interaction mode for neutrino energies below 1~GeV. Between 1 and several GeV the dominant mode is baryonic resonance production and decay, for instance through $\nu_\mu + p \to \mu^- + \Delta^{++} \to \mu^- + \pi^+ p$.

Sub-dominant contributions to the total cross-section come from meson exchange current (MEC) interactions (also referred to as the 2 particle-2 hole effect)~\cite{Martini:2009uj}, and from coherent and diffractive meson production. The MEC process requires the presence of two nucleons, where an electroweak boson from the leptonic current is exchanged by the nucleon pair, and is followed by 2-nucleon emission from the primary vertex. This generates pion-less final states similar to quasi-elastic scattering. This is particularly important for energies below 1~GeV. Finally, in coherent meson production the target nucleus remains in its ground state, leading to the production of a forward meson, for instance $\nu_\mu + \ce{^{16}O} \to \mu^- + \pi^+ + \ce{^{16}O}$. This is mainly important at low energies and momentum transfers. 

\begin{figure}[h] 
\centering
\includegraphics[width=0.49\textwidth]{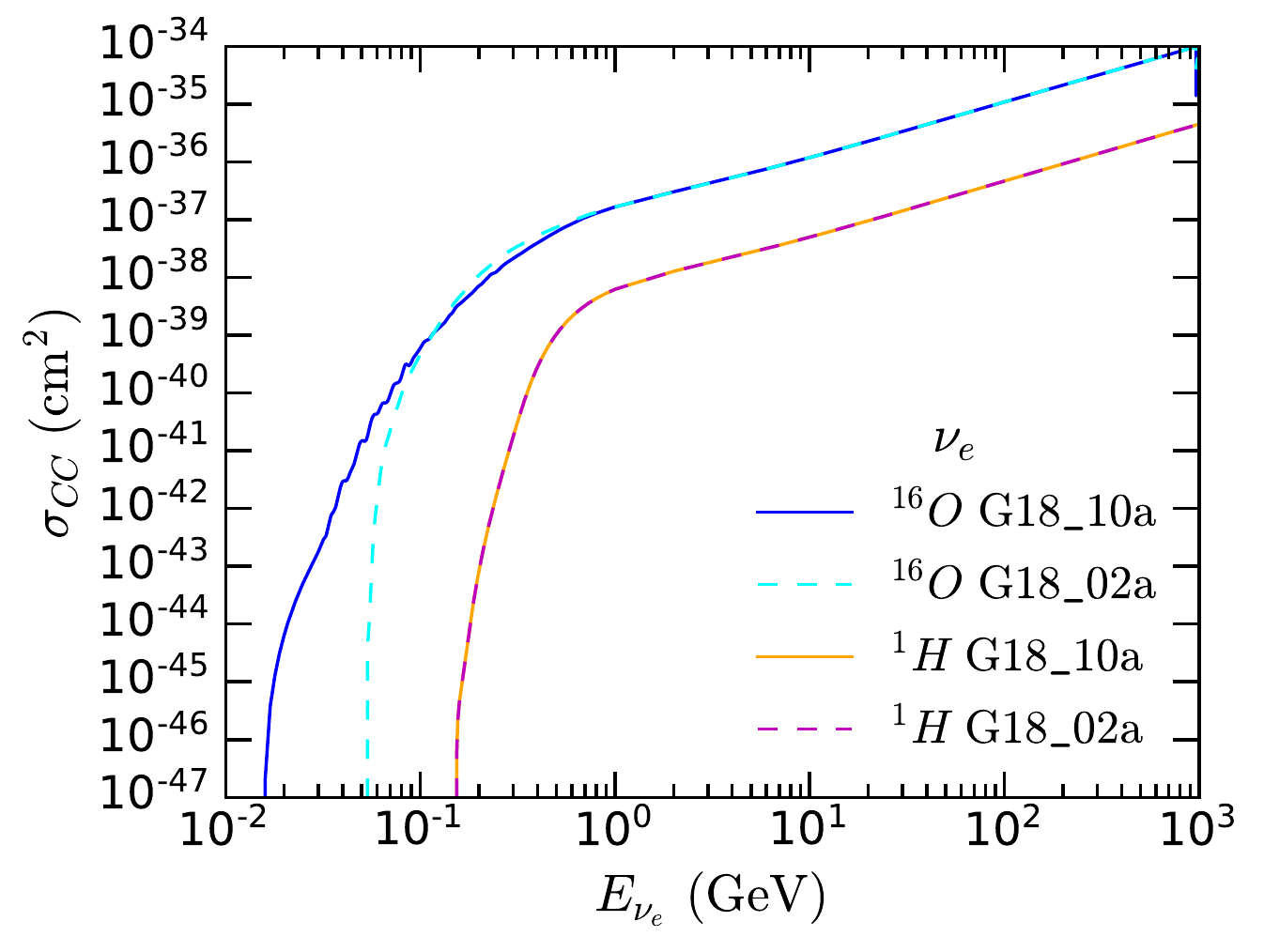}
\includegraphics[width=0.49\textwidth]{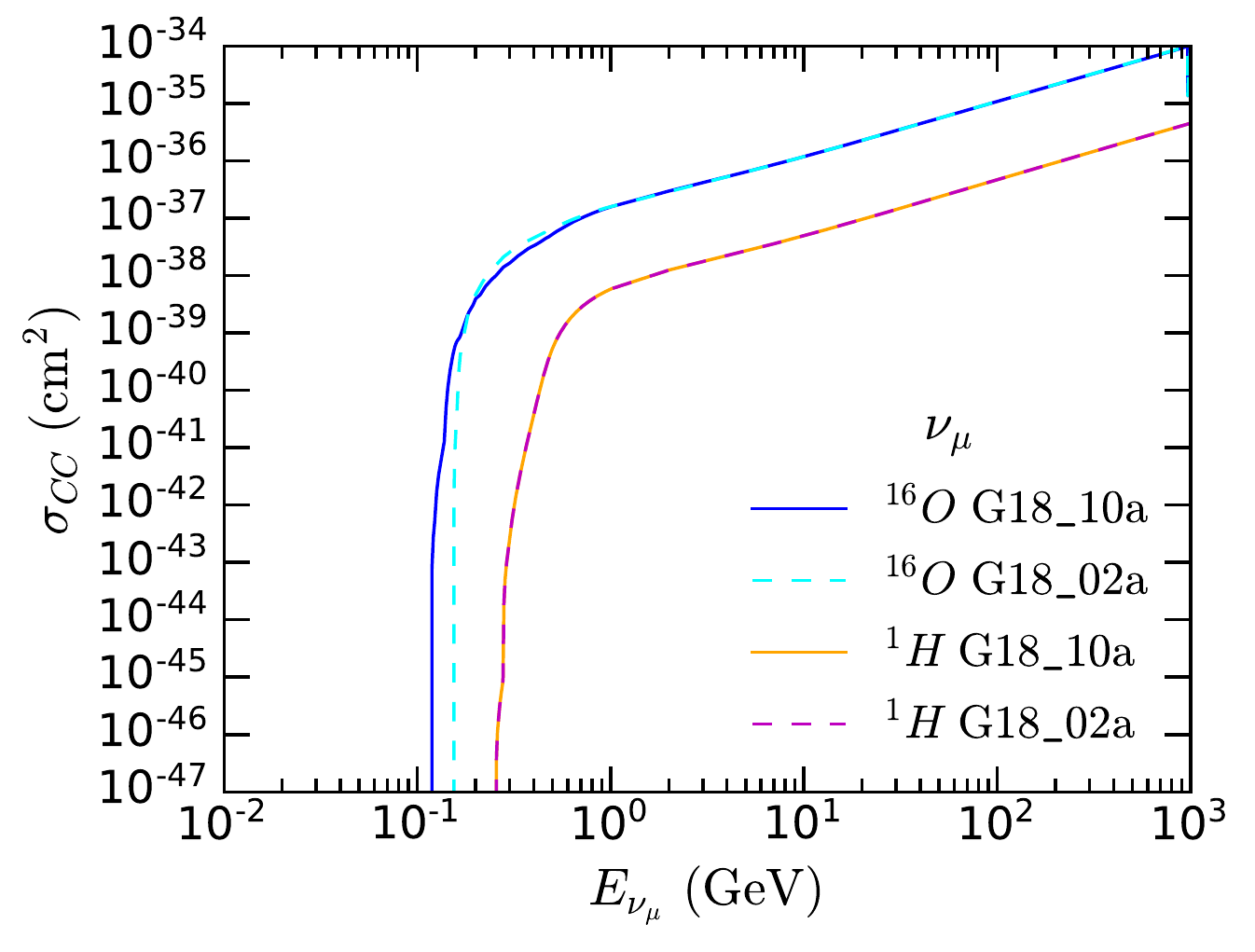}
\caption{The total charged current cross sections for $\nu_e$ (left) and $\nu_\mu$ (right) from ~\texttt{GENIE} pre-computed cross sections on hydrogen and oxygen, for tunes G18\_02a and G18\_10a. The main differences are at low energies for electron neutrinos.
}
\label{fig:xsec}
\end{figure}

All these interaction processes are modelled by \texttt{GENIE}. However, there are differences between the models used in the G18\_02a and G18\_10a tunes for the quasielastic and meson exchange current interactions. The G18\_02a tune uses the Llewellyn-Smith model~\cite{LlewellynSmith:1971uhs} for the QE interactions and an empirical MEC model, while the G18\_10a tune uses the Fermi-Gas approximation of~\cite{Nieves:2004wx,Nieves:2005,Nieves:2011pp} for both interactions. This leads to important differences in the interaction rates at low energies, particularly for electron neutrinos and anti-neutrinos. In particular, for energies of $\mathcal{O}(\rm{MeV})$, the cross-sections are larger in the G18\_02a tune by nearly a factor of two on oxygen nuclei. On the other hand below 100~MeV, the cross-sections are much larger for the G18\_10a tune. Accordingly, we study the effects of both tunes, with the  computed cross sections for $\nu_e$ and $\nu_\mu$ shown in Fig.\ref{fig:xsec}.

The SuperK and HyperK detectors do not measure the reaction modes of individual events. Instead, the detectors measure final states or topologies, such as final states with $1\mu^-$ and zero pions (denoted by $\{1\mu^-, 0\pi\}$). For example, quasi-elastic scattering mainly yields $\{1\mu^-, 0\pi\}$ events, with a background of misidentified charged current single pion production where the pion is absorbed or not seen by the detector. The final state produced by a given interaction determines the event category that an event is classified as. This is discussed in further detail in Section~\ref{sec:cat}, following an account of our tracking and smearing procedure in Section~\ref{sec:tracking}. First, however, we discuss the so-called invisible muon and spallation backgrounds.

Low energy atmospheric $\nu_\mu$ can interact to produce muons with $E\lesssim 50$~MeV, which is below the threshold for the muon to emit Cherenkov photons.  Consequently, when the muon decays, the resulting electron cannot be associated with the decay of a visible muon. These are known as invisible muons. These events form the dominant background in event categories that contain low energy $\nu_e$ and $\bar{\nu}_e$. The shape of the invisible muon background is determined by the Michel spectrum. We do not simulate the invisible muons in our detector simulation, but take the expected HyperK distribution (assuming neutron tagging with 70\% efficiency) between 10 and 50~MeV directly from Fig.~188 of the HyperK Design Report~\cite{Abe:2018uyc}. We then rescale these to correspond to 20 years of exposure. For the tail of the distribution above 50~MeV we use the same shape as in the SuperK DSNB search~\cite{Bays:2011si}.

Another important background at low energies is from muon-induced spallation products. These occur when muons from cosmic rays traverse through or near the detector and lead to the formation of radioactive isotopes. The decay products of these unstable isotopes are the dominant background for searches below 16~MeV~\cite{Abe:2018uyc}. The nature of these events at SuperK were the subject of a recent series of intensive theoretical studies for water Cherenkov detectors~\cite{Li:2014sea,Li:2015kpa,Li:2015lxa}, followed by SuperK observations~\cite{Super-Kamiokande:2015xra}. In the absence of details of these backgrounds in the HyperK DR, we consider 16~MeV as the lower threshold for our projected searches. Since HyperK is at a shallower site than SuperK, muon fluxes and hence spallation backgrounds are expected to be more important there. Accordingly, a detailed study of these backgrounds and their impact on dark matter searches at HyperK would be of great interest.

 Neutron tagging provides a means of mitigating both of these backgrounds. One method to achieve this is via the addition of gadolinium to the water in the detector at the 0.1\% level~\cite{Beacom:2003nk}, which will soon be implemented at SuperK. Gadolinium is a neutron absorber; following neutron capture, excited Gd nuclei de-excite through emission of 3-4 $\gamma$-rays with a total energy of $\sim8$~MeV after a characteristic time of 30~$\mu\rm{sec}$. Use of timing and vertex information can thus be used to tag neutrons, allowing the identification of true inverse beta-decay events, $\bar{\nu}_e + p \to n + e^+$,  providing a means of suppressing the invisible muon and spallation backgrounds. While Gd may also be used in HyperK, improved photosensors and photocathode coverage will enable a much higher efficiency for neutron tagging on {\it hydrogen} than is possible at SuperK. It is expected that the 2.2~MeV photon resulting from neutron capture on hydrogen will be detected in HyperK with an efficiency of order 70\%~\cite{Abe:2018uyc}. We do not model the neutron tagging process in our HyperK simulations, but instead take distributions for the invisible muon spectrum with and without neutron tagging from the HyperK Design Report.

\subsection{Tracking and smearing}
\label{sec:tracking}

Charged current interactions with neutrino initial states lead to the creation of charged leptons, which propagate in and through the detector volume. Accurately modelling the rate of energy loss of these particles is important for assigning events to the partially-contained or fully-contained event classes. The rate of energy loss (or mass stopping power) for heavy particles in matter is described by the Bethe-Bloch equation,
\begin{equation}
    \left\langle - \frac{dE}{dx}  \right\rangle = a(E) + b(E)E \,, 
\end{equation}
where $a(E)$ and $b(E)$ represent the electronic stopping power and losses due to radiative processes, respectively. We use this formula to model the track length of muons in liquid water and ``standard rock" (with Z=11 and A=22) using an interpolating function based on the energy-dependent parameters from~\cite{Groom:2001kq}\footnote{Available online at \href{http://pdg.lbl.gov/AtomicNuclearProperties/}{pdg.lbl.gov/AtomicNuclearProperties/}}. The muon energy loss rate is dominated by ionisation for muon energies less than about 100~GeV, above which radiative and other losses become important. For electrons, we take the radiation lengths from the tables in ref.~\cite{Tsai:1973py} and the critical energy from the ``Passage of particles through matter" section of the PDG~\cite{Tanabashi:2018oca}. The lepton energies and momenta are given  for each event by \texttt{GENIE}, as described above.

\begin{figure}[h]
    \centering
    \includegraphics[width=0.495\textwidth]{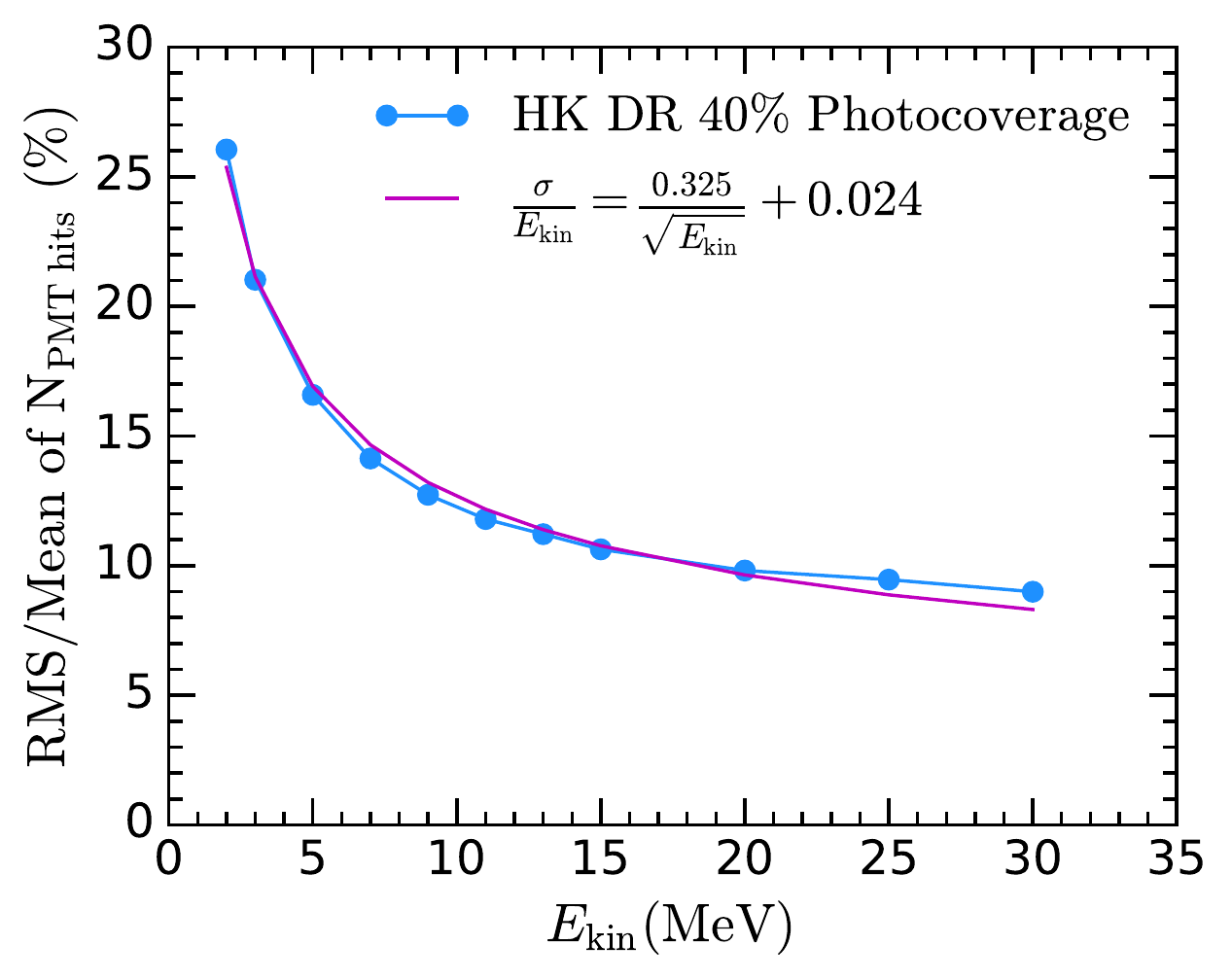}
    \includegraphics[width=0.49\textwidth]{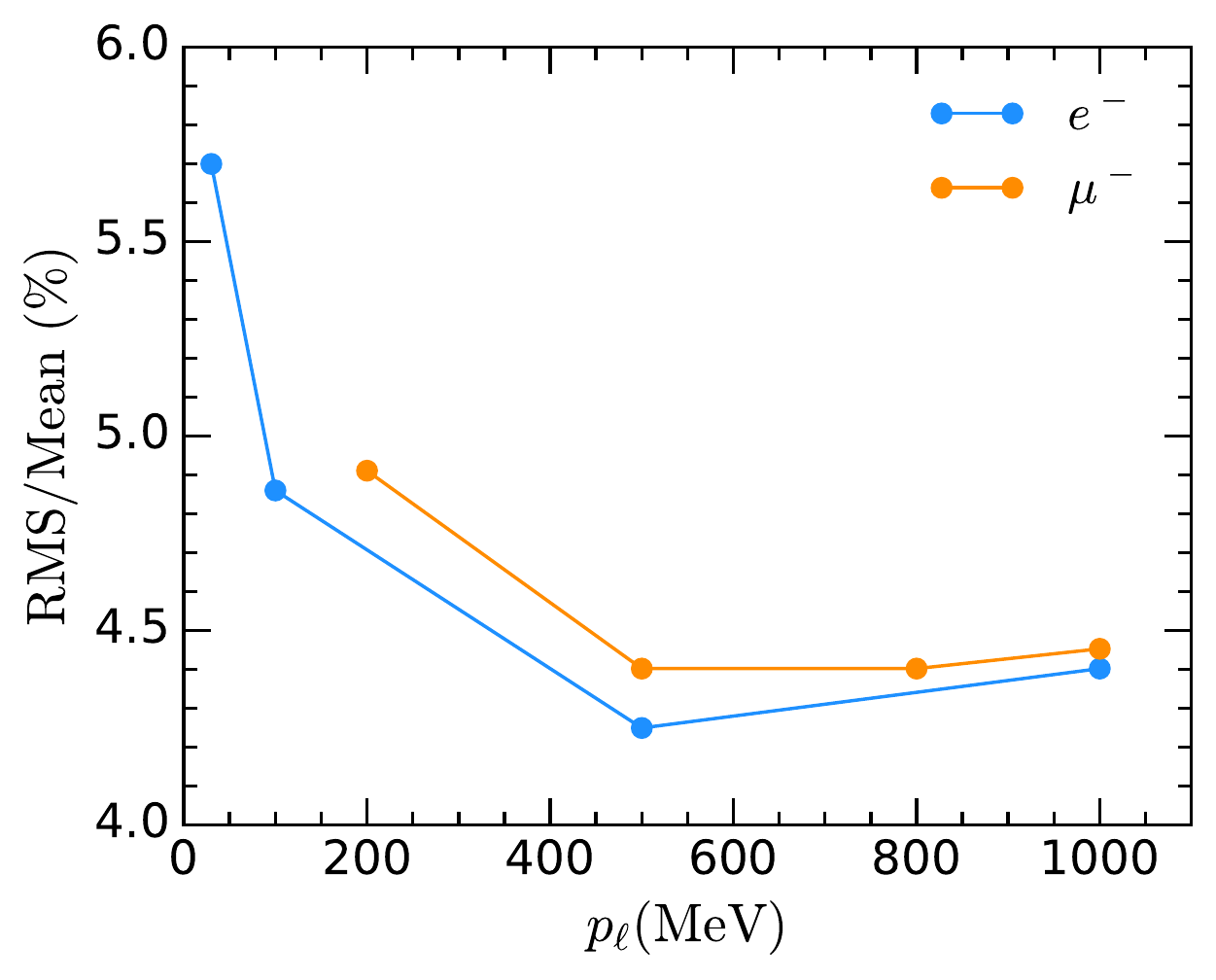}    
    \caption{Left: RMS/mean of the number of PMT that register a hit. We show data taken from Fig.~113 of the HyperK DR~\cite{Abe:2018uyc} for electrons injected with different  kinetic energy values assuming 40\% photocoverage (light blue points and line), together with our energy resolution fit (magenta). 
    Right: RMS/total charge distributions for electrons (light blue) and muons (orange).}
    \label{fig:energyres}
\end{figure}

\begin{figure}
    \centering
    \includegraphics[width=0.49\textwidth]{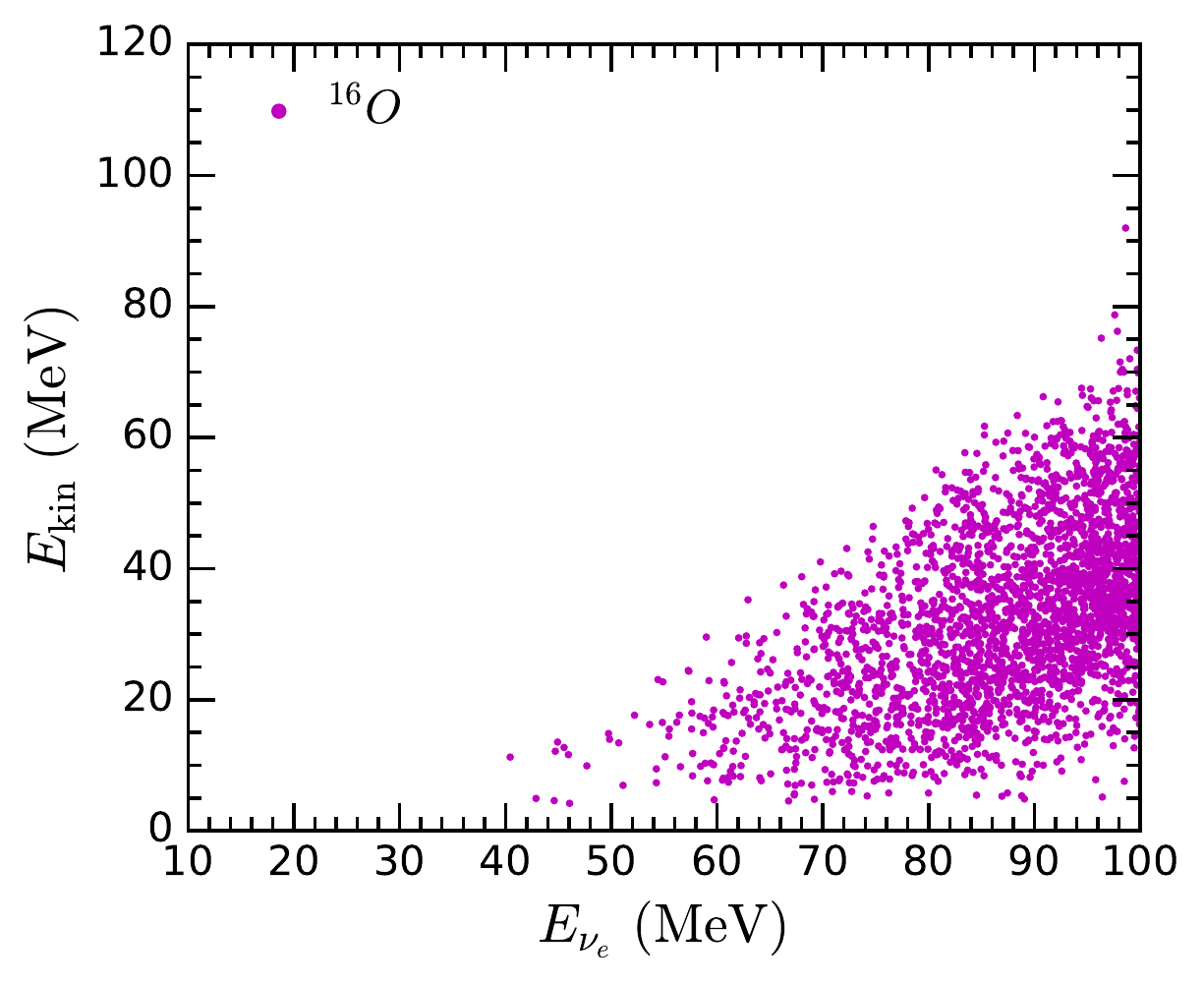}
    \includegraphics[width=0.49\textwidth]{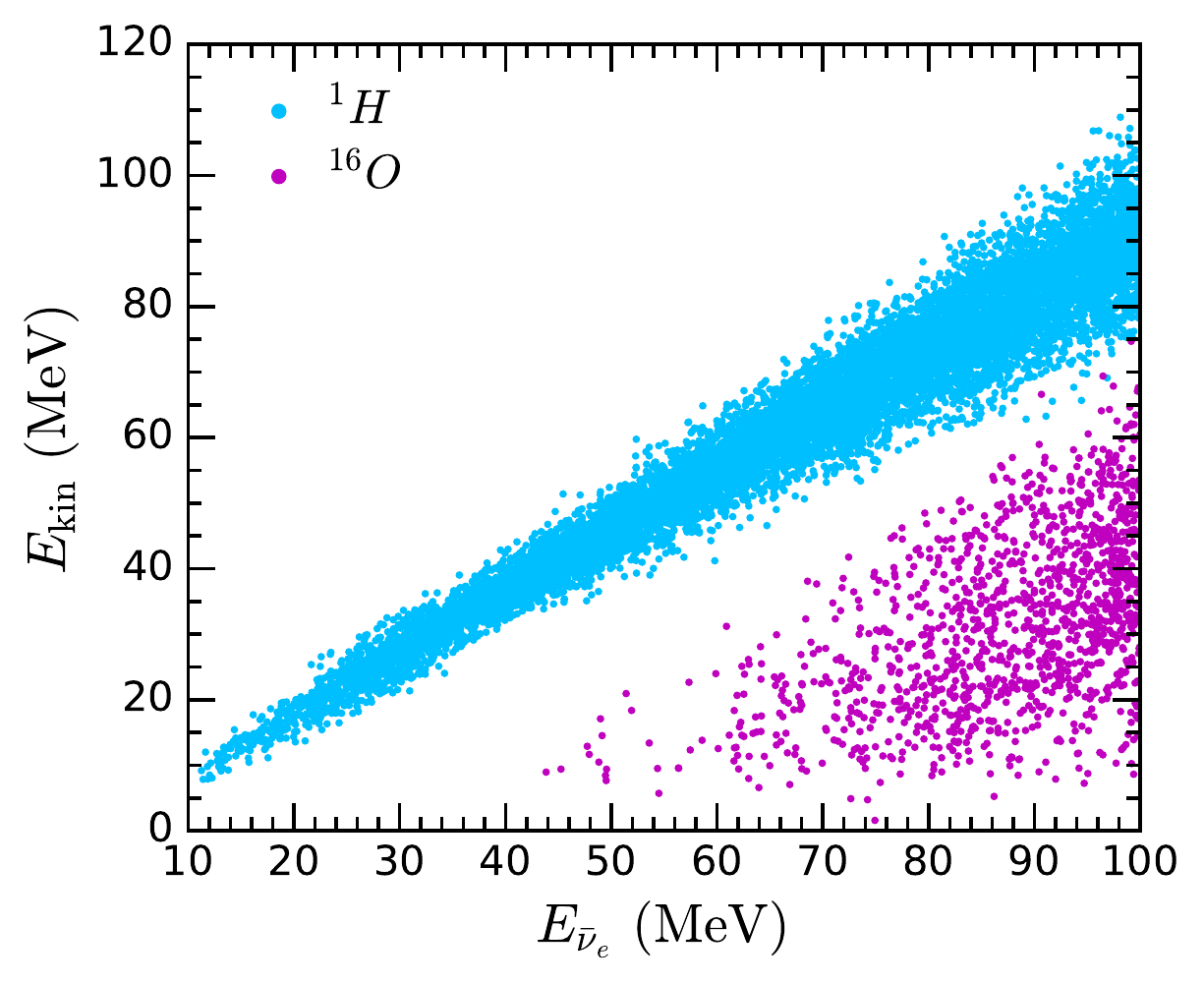}    
    \caption{$\Ekin$ vs. neutrino energy $E_{\nu}$ for $\nue$ (left) and $\nuebar$ (right) events in Fig.~\ref{fig:FCnueFLUKA_Ekin}. Magenta dots correspond to scattering off oxygen, and blue dots to scattering off hydrogen. The neutrino scattering cross-section off hydrogen is suppressed, as shown in Fig.~\ref{fig:xsec}.}
    \label{fig:EvsEkin}
\end{figure}

The energy resolution of the HK detector is determined by  the photocathode coverage. 
For low energy particles with $\Ekin\leq30~\MeV$, the resolution can be approximated using the number of PMT hits. The number of PMT hits is related to $E_{\rm{kin}}$, and we use an interpolating formula based on the  PMT hit distribution given in Fig.~113 of ref.~\cite{Abe:2018uyc}. We find that an adequate fit can be obtained using the same functional form used for the SuperK resolution in ref.~\cite{PalomaresRuiz:2007eu}. This gives the energy resolution in MeV as 
\begin{equation}
\sigma = 0.325\sqrt{\Ekin/\MeV}+0.024\Ekin,    
\end{equation}
where $\Ekin$ is given in MeV. A comparison with the Design Report results is shown in the left panel of Fig.~\ref{fig:energyres}. For higher energy leptons with $\Ekin>30\MeV$ we use the total charge distribution at several electron and muon momenta shown in the right  panel of Fig.~\ref{fig:energyres}, which is adapted from Fig.~112 of ref.~\cite{Abe:2018uyc}, again assuming 40\% PMT coverage. In this case we use a linear spline interpolation. For leptons with momentum higher than 1 GeV, we keep the energy resolution of a lepton with $p_\ell=1\GeV$. The kinetic energy, $\Ekin$, of each event will be then smeared by a Gaussian distribution with a mean $\Ekin$ and width $\sigma(\Ekin)$. Note that we present our distributions in terms of the final state lepton kinetic energy, $\Ekin$. This is not directly measured by SuperK. Rather, they make cuts on a quantity $\Evis$, which is the energy of an electromagnetic shower that yields the same amount of Cherenkov light~\cite{Jiang:2019xwn} in a given event. We show the relation between $E_\nu$ and $\Ekin$ for electron neutrinos and anti-neutrinos at low energies in the left and right-hand panels of Fig.~\ref{fig:EvsEkin}, respectively. The diffuse magenta dots in the bottom left are from scattering off oxygen, while the more tightly grouped blue dots in the right panel correspond to scattering off hydrogen. No hydrogen scattering is visible for the neutrinos in the left-hand panel since the cross-section is highly suppressed (see Fig.~\ref{fig:xsec} left).

In principle one could simulate the individual PMT responses using a code such as \texttt{WCSim}~\cite{wcsim}, thus allowing calculations and cuts based on $E_{\rm{vis}}$. We also do not smear the location of the primary vertex. This makes only a small difference to events that are located near the boundary of the fiducial volume.

\subsection{Event Categories}
\label{sec:cat}

 Depending on where the interaction takes place, the direction of the resulting final state products, and how they appear in the detector, SuperK and HyperK divide events into a number of different classes. For validating our detector simulation and projecting limits for DM indirect detection at HyperK, we will consider the fully contained (FC) and partially contained (PC) categories, for electron and muon neutrinos and anti-neutrinos. Fully contained events are those in which all the energy is deposited in the inner detector. A partially contained event involves a high-energy muon that leaves the inner detector and deposits energy in the outer detector.
 A third category used in atmospheric neutrino and other DM analyses involves muons created in the rock surrounding the detector and then observed travelling up through the detector volume: these are upward-going muons (Up-$\mu$). While important at higher energies, at the low neutrino energies we are studying the number of upward-going muons are small~\cite{Ashie:2005ik}, and so we omit them from our study. Downward-going muons samples are highly contaminated by cosmic rays and so not used.

FC and PC events can be subdivided into further categories, based on the properties of the Cherenkov rings, such as their number, energy and particle ID. We do not simulate the Cherenkov radiation, instead keeping all the FC events in a single class (and similarly for the PC events), and distinguishing only between electron and muon neutrinos. Identification of $\mu^{\pm}$ and $e^\pm$ can be done with high efficiency based on the properties of the associated Cherenkov rings. On the other hand, SuperK and HyperK cannot distinguish the charge of the particle responsible for a given ring. There are likelihood based methods that enable the construction of event samples that are enriched in $\nue$ and $\nuebar$ respectively, but these are far from pure and only apply to Multi-GeV ($E_{\rm{vis}}>1.33$~GeV) events, not the Sub-GeV events we study. Thus, our analysis will be based on three composite event classes: FC$(\nue+\nuebar)$, FC$(\numu+\numubar)$ and PC$(\numu+\numubar)$.

 Fully contained events are generated inside the inner detector (ID), with all secondary particles also required to be within the ID. The primary vertex must lie within the fiducial volume, which is the region of the ID with a boundary 1.5~m inside the inner wall~\cite{Abe:2018uyc}. A lower energy cut similar to that applied to SK events \cite{Ashie:2005ik}, $P_\mu>200$ MeV for $\numu$ and $\numubar$ events, has been imposed on single-ring events \cite{Ashie:2005ik}, along with a cut $\Ekin>30$~MeV. We do not apply such a cut to $\nue$ and $\nuebar$ events.
 Multi-ring events are induced if the neutrino interaction gives rise to further charged particles (e.g. $\pi^\pm$) along with the resulting lepton. In ref.~\cite{Ashie:2005ik}, multi-ring events were identified as $\mu$-like when the most energetic ring had $P_\mu > 600$ MeV and visible energy  $E_{\rm vis} > 600$ MeV. Multi-ring events are not an important contribution in our study. At low neutrino energies the dominant interaction modes are charged-current quasi-elastic scattering (CCQE) and meson-exchange current (MEC) interactions. Both of these lead to single-ring events. 
 
 Partially contained events are those with a primary vertex within the inner detector but with a high energy muon exiting the ID fiducial volume. These events are further required to have a minimum track length of 2.5 m and muon momentum $P_\mu \geq 700$~MeV. PC events are thus generally associated with higher energy events, and are less important than FC events for deriving limits on light dark matter.
 
We close this section with a comment on triggers. The current SuperK searches for dark matter use the atmospheric neutrino event samples based on the High Energy (HE) trigger. Lower energy searches, such as for the DSNB, use a separate Low Energy (LE) trigger. HyperK will have a similar trigger structure (there is a also a Super Low Energy trigger relevant for solar neutrino analyses). This use of the atmospheric neutrino event samples is part of the reason that SuperK have not extended their DM limits to lower masses. The HyperK DR is similar, and in fact the DM Monte-Carlo event sample used there is simply a sample of reweighted atmospheric neutrinos. Setting limits on DM annihilation from low energies up to 1~GeV may require the use of the LE trigger, or combining LE- and HE-based event samples in some way.

\subsection{Validation}
\label{sec:validation}

\begin{figure}[h] 
\centering
\includegraphics[width=\textwidth]{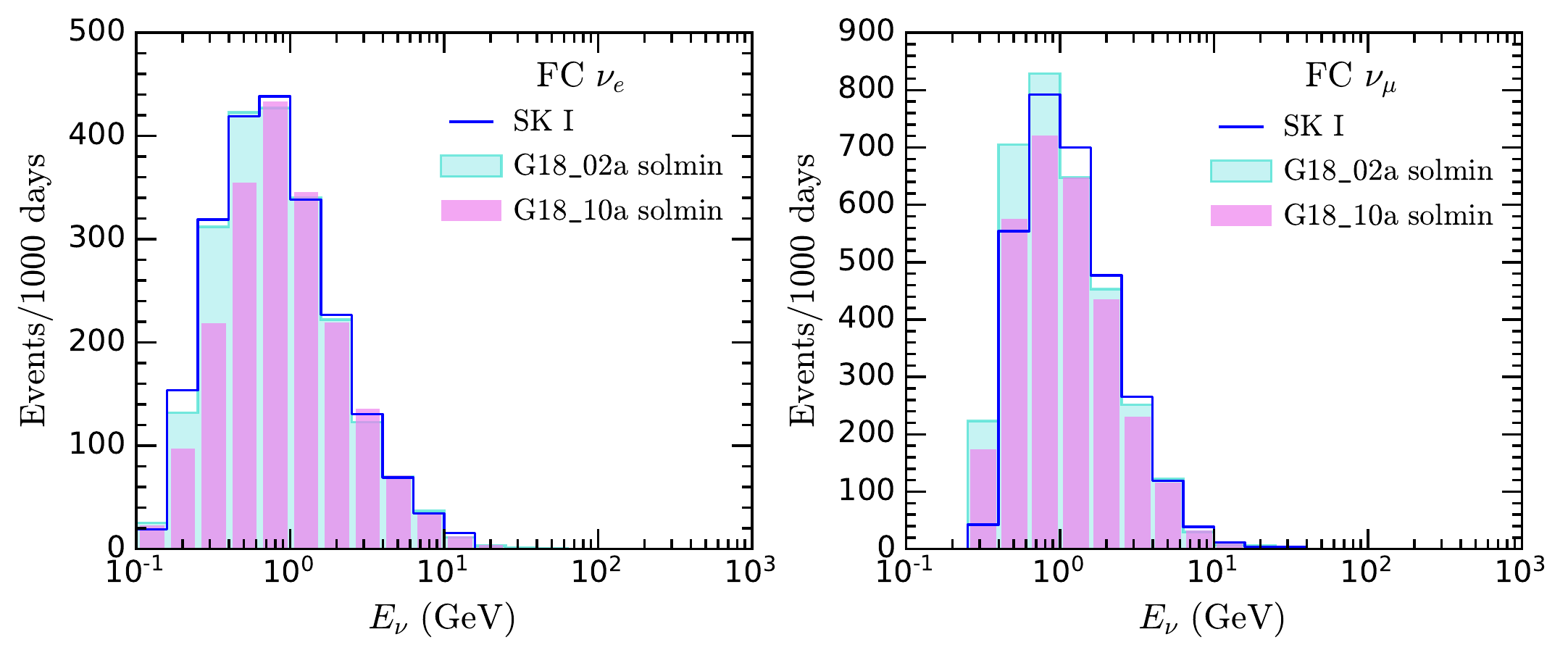}
\caption{FC $\nue$ (left) and $\numu$  (right) event rate for SK. Electron neutrino events selected by reaction mode and muon neutrinos selected by topology. Results are shown for the G18\_02a and G18\_10a \GENIE tunes, and compared with the expected events for SK I taken from Fig.~1 of ref.~\cite{Ashie:2005ik}, as labelled.\label{fig:FCvalidation}
}
\end{figure}

\begin{figure}[h] 
\centering
\includegraphics[width=0.49\textwidth]{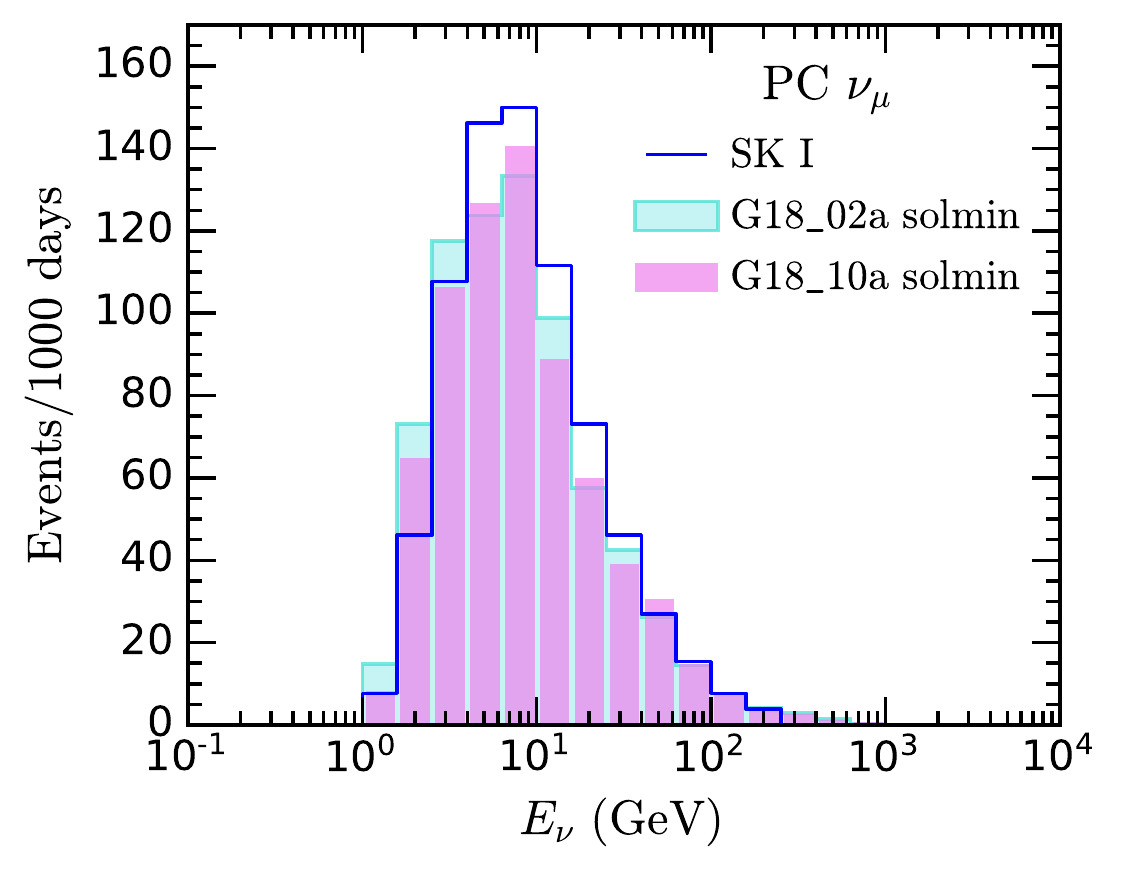}
\caption{PC $\numu$ event rate for SK. Events selected by topology. 
Results are shown for the G18\_02a and G18\_10a \GENIE tunes, and compared with the expected events for SK I taken from Fig.~1 of ref.~\cite{Ashie:2005ik}. 
 \label{fig:PCvalidation}
}
\end{figure}

We have validated our detector simulation and analysis pipeline against SuperK measurements of the atmospheric neutrino spectrum~\cite{Ashie:2005ik}.

There are more up-to-date atmospheric neutrino measurements than~\cite{Ashie:2005ik} by SuperK, but they incorporate data from multiple running configurations of the experiment, which have substantial differences between them. These include decreased PMT coverage in SK-II (due to the implosion incident in 2001) and the improvements in reconstruction electronics in SK-IV. Rather than attempt to reproduce all of these, we choose instead to validate our simulations against data from SK-I, and then scale our simulation up to the dimensions and performance parameters of HyperK.

The SuperK measurements took place over three years of solar minimum, one transition year and a single year of solar maximum for a total of 1489 live-days of data. We compare this with simulations for the flux at solar minimum.
We show in Fig.~\ref{fig:FCvalidation}  the differential event-rate per 1000 days as a comparison between our results for fully contained electron (left) and muon (right) neutrinos, compared with the data from Fig.~1 of~\cite{Ashie:2005ik} shown as a solid blue line. We show both the 
G18\_02a and G18\_10a tunes in cyan and magenta respectively. Fig.~\ref{fig:PCvalidation} is a similar plot for the partially contained muon category. At the lowest energies $E_\nu \lesssim 100$~MeV the scattering cross-section for the G18\_02a tune is much smaller than for G18\_10a. However, between 100~MeV and 1 GeV  the G18\_02a cross-section is slightly larger, as can be seen in Fig.~\ref{fig:xsec}. We see that our simulation matches the SuperK data quite well. We have examined selecting events in categories by either specific reaction modes and by final state topology. For $\nue$ events we find that selecting by reaction mode gives better agreement with the SK data, while for $\numu$ we find that selection by topology does. Consequently, we adopt these slightly different criteria for $\nue$ and $\numu$ for the rest of our study.

We have also compared the acceptance of our HyperK simulation with that of our SuperK simulation. While broadly similar there are some changes explained by the differing geometric sizes of the detectors. For instance, we find that more events at higher energy make it into the fully contained categories at HyperK than at SuperK. We use the G18\_10a in our HyperK analysis, since it is more up to date and the prescriptions for the CCQE and other cross-sections are more similar to the latest version of \texttt{NEUT}~\cite{Hayato:2002sd,Abe:2017aap,Jiang:2019xwn} used by SuperK and HyperK.

\begin{figure}
    \centering
    \includegraphics[width=\textwidth]{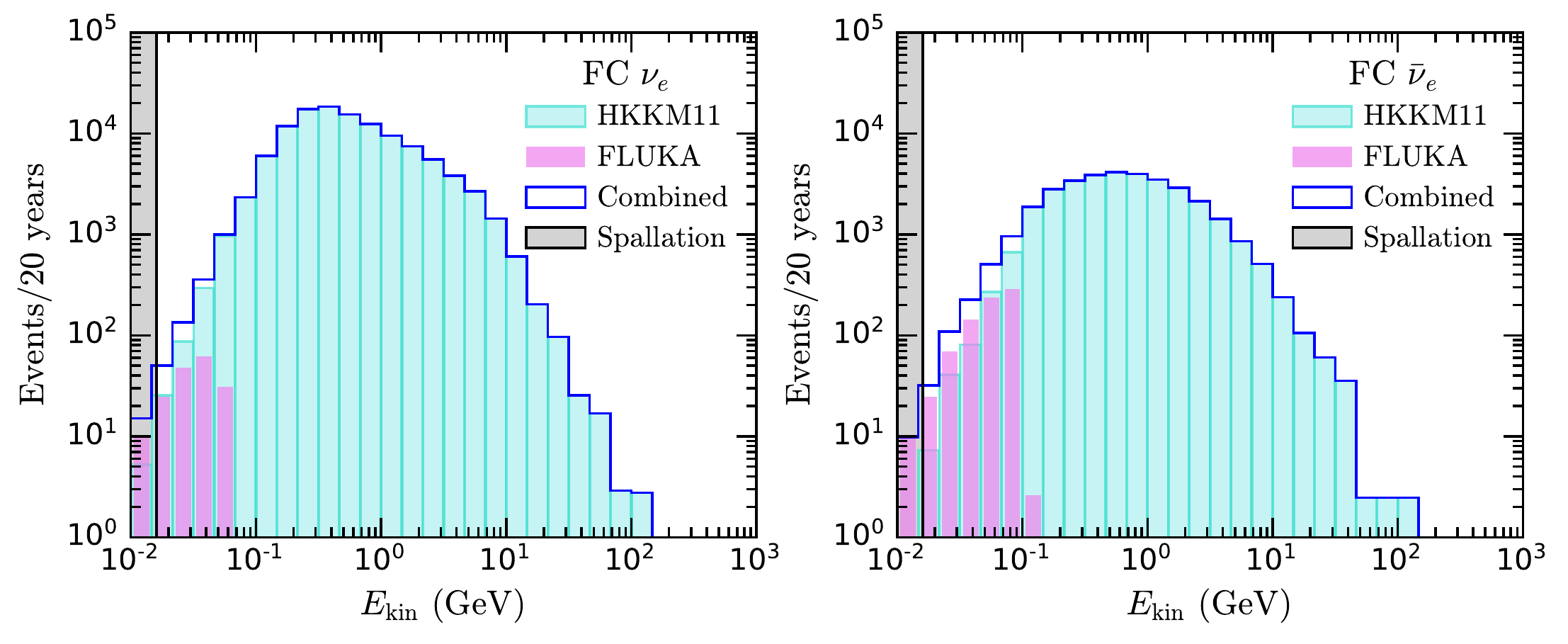}
    \caption{Combined fully contained $\nue$ (left) and $\nuebar$ (right) event rates at HyperK for the HKKM11 and FLUKA (below 100 MeV) fluxes, calculated with \GENIE tune G18\_10a and 20 years of livetime.\label{fig:HKbg} }
\end{figure}

\begin{figure}
    \centering
    \includegraphics[width=0.55\textwidth]{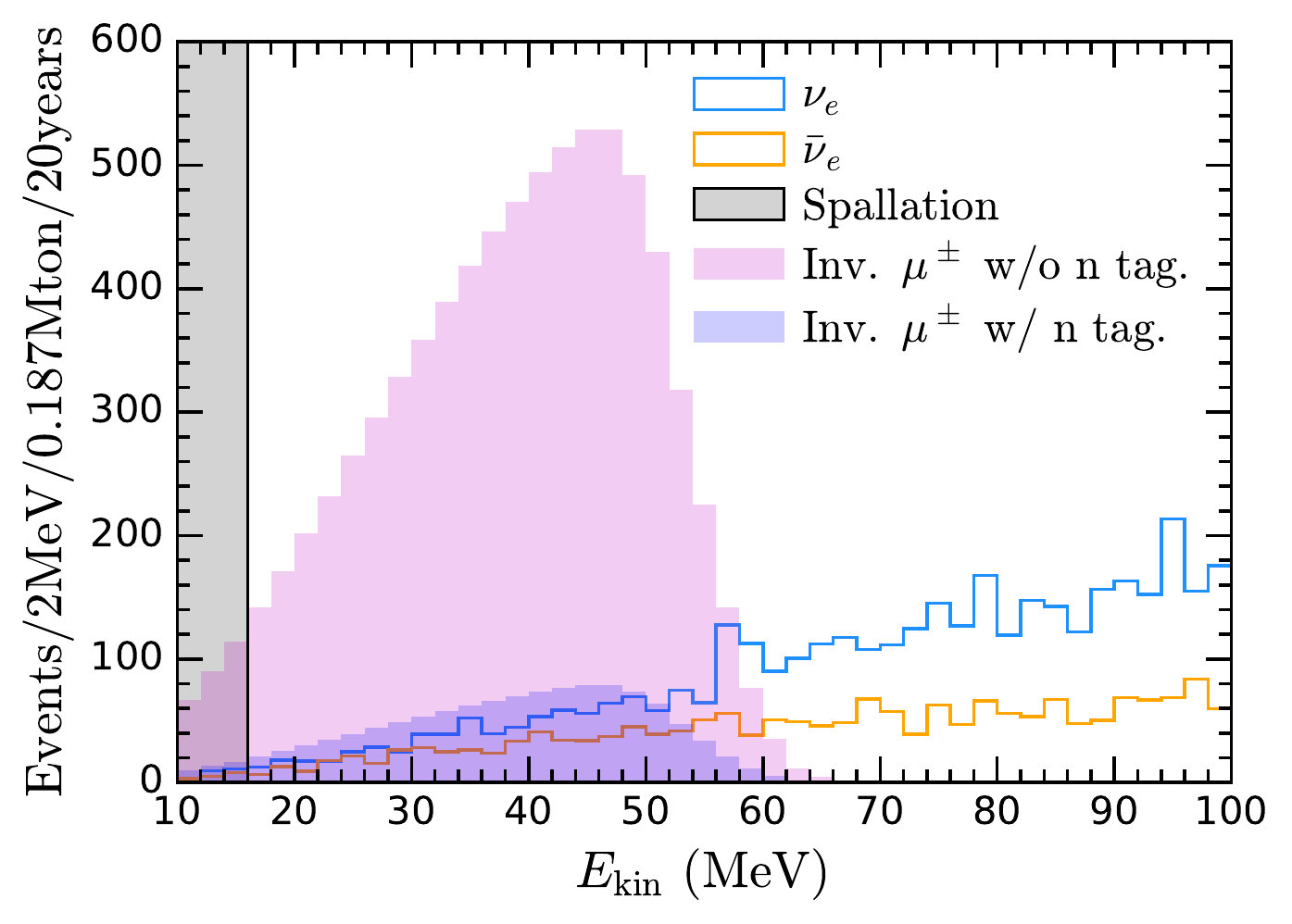}
    \caption{The expected $\nue$ (upper solid line, blue) and $\nuebar$ (lower solid line, orange) event rates below 100 MeV (obtained from FLUKA +  HKKM11) at HyperK  as a function of the final lepton kinetic energy, $\Ekin$. The coloured regions show the background contributions from invisible muons with (blue) and without (magenta)  neutron tagging, assuming a 70\% tagging efficiency. The grey area at the left is dominated by spallation backgrounds.}
     \label{fig:FCnueFLUKA_Ekin}   
\end{figure}

In Fig.~\ref{fig:HKbg} we show the total atmospheric backgrounds we use for the FC $\nue$ (left) and $\nuebar$ (right) categories, given by the sum of the HKKM11 and FLUKA fluxes. The grey band at the left hand side of the plots shows where the spallation backgrounds become dominant. Fig.~\ref{fig:FCnueFLUKA_Ekin} shows the low energy region  and the invisible muon backgrounds, with and without neutron tagging. Both sets of plots are for 20 years of running time. HKKM11 events were generated for 10 years of solar minimum and 10 years of solar maximum, while FLUKA events are for 20 years of solar average.
In our final analysis we sum over these backgrounds since HyperK cannot effectively discriminate between $\nue$ and $\nuebar$.

\section{Dark Matter Search and Projected Limits}
\label{sec:limits}

We now provide details of our calculation of the dark matter signal from the Galactic Centre. 
The differential flux of neutrinos from dark matter annihilation is given by 
\begin{equation}
\frac{{d\Phi_\nu}_{\Delta\Omega}}{dE_\nu} = \frac{\sigmav}{8\pi \mdm^2}  \Javg \frac{dN_\nu}{dE_\nu} \, ,
\label{eq:GCneuflux}
\end{equation}
where $\mdm$ is the dark matter mass, $\sigmav$ is the thermally averaged annihilation cross-section,  $\Javg$ is the angle-averaged J-factor defined below, and $\frac{dN_\nu}{dE_\nu}$ is the differential neutrino energy spectrum (obtained from \texttt{DarkSUSY}). This equation holds for the sum of the neutrino and anti-neutrino fluxes, and for Majorana dark matter. For the neutrino final state the differential flux at production is a delta function (although this becomes smeared by detector effects), for muons it becomes smeared out due to the muon decays.

The J-factor for DM annihilation in the ($b$, $l$) direction in Galactic coordinates is obtained integrating the DM density squared over the line of sight $s$~\cite{Gordon:2013vta}, 
\begin{equation}
J(b,l) = \int_0^{\smax} \rho^2\left(\sqrt{\rsun^2-2s \, \rsun\cos b \cos l+s^2}\right) ds, 
\end{equation}
where $\rsun=8.5\kpc$ is the distance from the Solar System to the Galactic Centre and
\begin{equation}
\smax = \sqrt{\RMW^2-\rsun^2+\rsun^2 \cos^2 b \cos^2 l} + \rsun \cos b \cos l,
\end{equation}
where $\RMW=40\kpc$ is the Galaxy halo size. The J factor averaged over a solid angle $\Delta\Omega$ is then defined as 
\begin{equation}
 \Javg =  \frac{1}{\Delta \Omega}\int_{\Delta \Omega} J(b,l)  \,  d\Omega,
\end{equation}
where $d\Omega = \cos b \, db \, dl$.

To quantify some of the astrophysical uncertainties involved in our limits we will present results for  three different dark matter halo profiles, which can all be expressed in the form
\begin{equation}
\rho(r) = \frac{\rho_0}{\left(\frac{r}{r_s}\right)^\gamma\left[1+\left(\frac{r}{r_s} \right)^\alpha \right]^{(\beta-\gamma)/\alpha}} \, .  \label{eq:DMdensprofile}
\end{equation}
These are the standard Navarro-Frenk-White (NFW) profile~\cite{Navarro:1995iw}, as well as the Moore~\cite{Moore:1999gc}, and Isothermal~\cite{Bahcall:1980} profiles. The appropriate values of alpha, beta and gamma for the chosen profiles can be found in Table~\ref{tab:DMprofiles}. The Moore profile is cuspier than NFW and leads to a larger J-factor, and hence a larger DM signal and stronger limits. The Isothermal profile is less cuspy and leads to weaker limits than NFW. The different choices in $r_s$ and $\rho(\rsun)$ are from SuperK and the HyperK Design Report.

\begin{table}[h]
    \centering
    \begin{tabular}{|l|c|c|c|c|c|}
    \hline
       Halo model & $\alpha$ & $\beta$ & $\gamma$ & $r_s[\kpc]$ & $\rho(\rsun)[\GeV\cm^{-3}]$\\
      \hline 
       NFW  & 1 & 3 & 1 & 20 & 0.3\\
       Moore & 1.5 & 3 & 1.5 & 28 & 0.27\\
       Isothermal & 2 & 2 & 0 & 5 & 0.3\\
       \hline
    \end{tabular}
    \caption{Parameters of eq.~\ref{eq:DMdensprofile} that determine the NFW~\cite{Navarro:1995iw}, Moore~\cite{Moore:1999gc} and Isothermal~\cite{Bahcall:1980} density profiles.
    }
    \label{tab:DMprofiles}
\end{table}

For simplicity we have fixed the location of the GC at the zenith  $z=90^\circ$ of the HyperK experiment. We have investigated the effects of changing the GC location relative to the detector orientation but find them to be relatively small (see appendix~\ref{sec:app2}), since HyperK has approximately the same measurements in all dimensions.  We also do not include the contribution from extragalactic dark matter, which leads to a diffuse and isotropic signal. This has been used in the past to set conservative limits on the total DM annihilation cross-section~\cite{Beacom:2006tt}. The HyperK DR does not include the extragalactic contribution in setting their limits, and so we follow them in order to achieve a fair comparison with their results.  Unlike the GC contribution, whose spectrum has the form of a delta-function, the extragalactic neutrino spectrum is smeared out by the effects of redshift. The total flux however is of similar size to that from the GC~\cite{Klop:2018ltd}. 
Consequently, one may think of our analysis as a conservative all-sky analysis that uses signal only from DM annihilations in the Galactic Centre, but includes all relevant background contributions.
Accordingly, including the extragalactic contribution would lead to slightly better limits, as would binning in $\cos\theta_{\rm GC}$, see ref.~\cite{Arguelles:2019ouk} for further recent discussion.

The strictly correct way to do the coordinate transformation between Galactic and horizon coordinates is from Galactic to equatorial coordinates (centred at Earth) and then to horizon coordinates at the detector location.
We use a simplifying assumption for the coordinate transformation relations between Galactic and horizon coordinate systems, assuming that the GC is located at azimuth angle $0^\circ$ and zenith angle $\zGC$, is obtained by rotating the $z$ axis (north Galactic pole) by  $90^\circ-\zGC$, 
\begin{eqnarray}
\begin{bmatrix}
   \cos a \sin z \\
   \sin a \sin z\\
   \cos z
  \end{bmatrix}&=& R_y\left(\frac{\pi}{2}-\zGC\right)
\begin{bmatrix}  
   \cos l \cos b \\
   \sin l \cos b\\
   \sin b  
\end{bmatrix}  \\
&=& \begin{bmatrix}  
\sin \zGC & 0 &  -\cos \zGC \\
0 & 1 & 0 \\
\cos \zGC & 0 & \sin \zGC
\end{bmatrix}  
\begin{bmatrix}  
   \cos l \cos b \\
   \sin l \cos b\\
   \sin b  
\end{bmatrix}.
\end{eqnarray}
The transformation from $(l,b)$ to $(a,z)$ is given by, 
\begin{eqnarray}
\tan a &=& \frac{\sin l}{\sin \zGC \cos l- \cos \zGC \tan b},\\
\cos z &=& \sin \zGC \sin b + \cos \zGC \cos b \cos l .
\end{eqnarray}

\begin{figure}[t]
\centering
\includegraphics[width=0.49\textwidth]{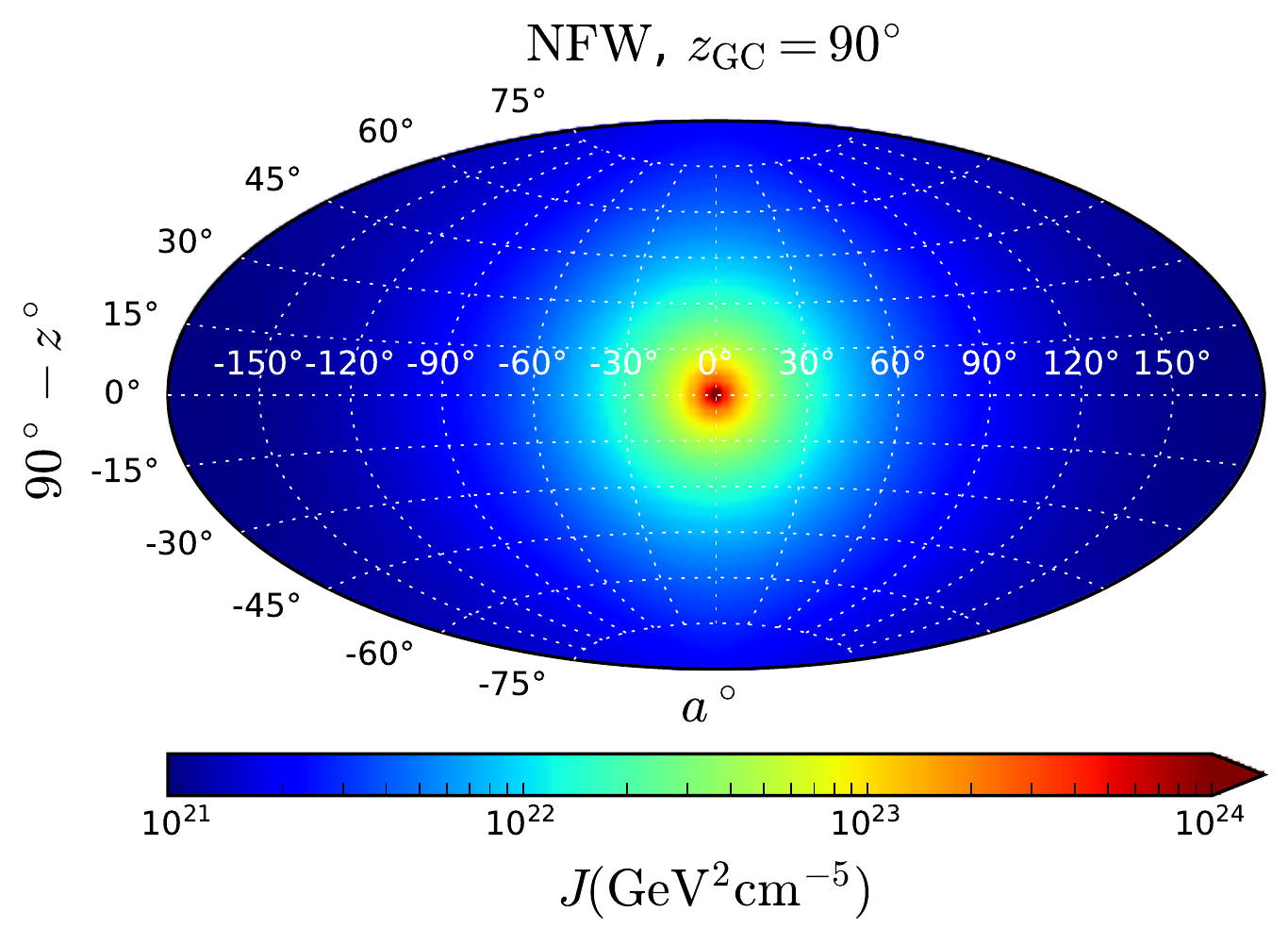}
\includegraphics[width=0.49\textwidth]{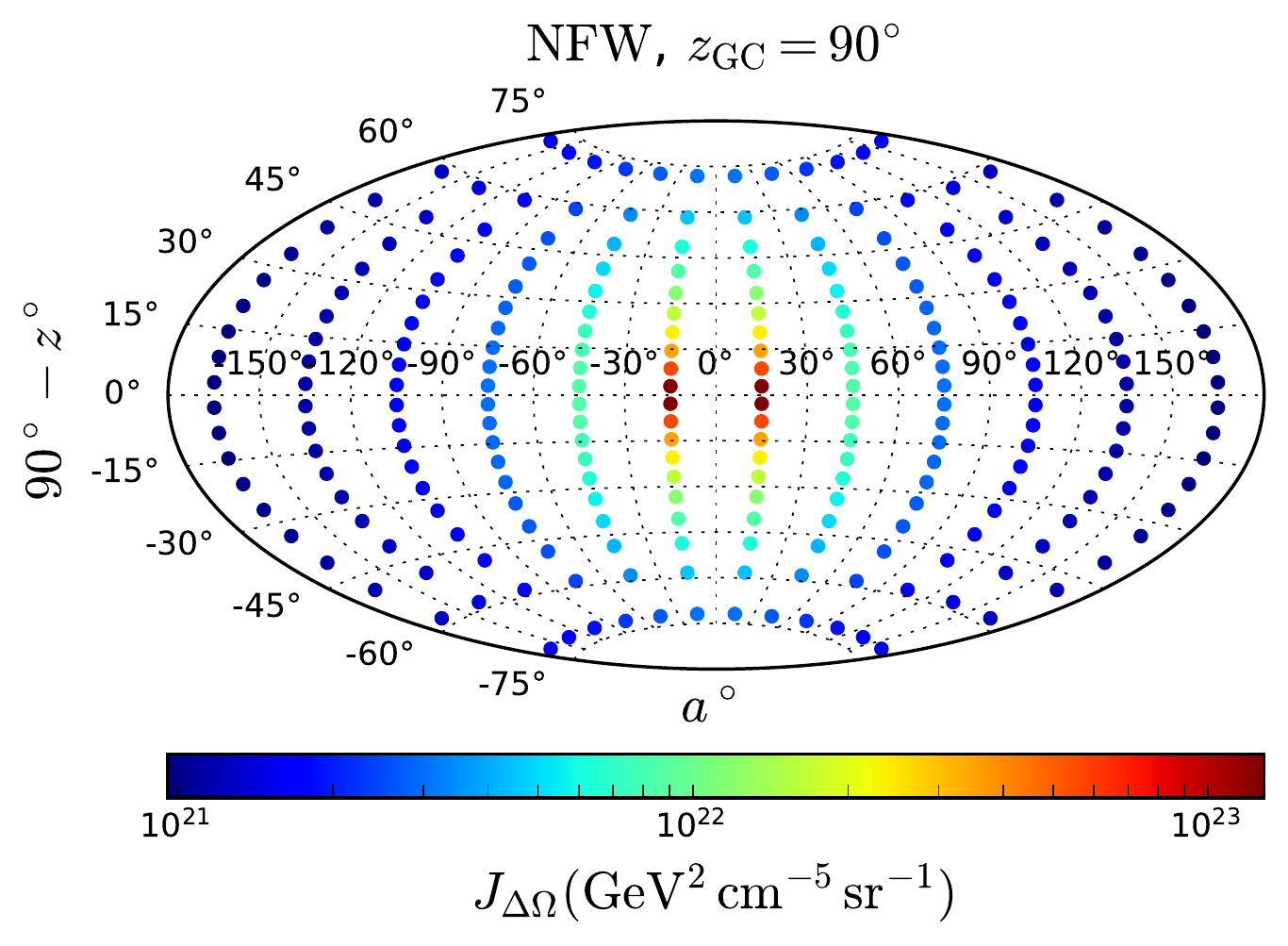}
\caption{Left: The J-factor in horizontal coordinates, assuming that the GC is located at altitude $0^\circ$ and azimuth $a=0^\circ$. 
Right: Same as left but binned in solid angle $\Delta\Omega$ as described in the text.}
\label{fig:Jfactorbinned}
\end{figure}
We bin the J-factor in the following way.
As for the background  HKKM11 flux, we have considered 12 bins in azimuth and 20 bins in $\cos z$, 
\begin{eqnarray}
\Delta\Omega_{i,j} &=& \Delta a_i \int_{z_j}^{z_{j+1}} \sin z \, dz, \\
J_{\Delta\Omega_{i,j}} &=&  \frac{1}{\Delta \Omega_{i,j}} \int_{a_i}^{a_{i+1}} \int_{z_j}^{z_{j+1}} J(a,z)  \, \sin z \,  dz \, da.
\label{eq:Jfactorbinned}
\end{eqnarray}
The J-factor binned in solid angle is shown in the right panel of Fig.~\ref{fig:Jfactorbinned}, where we have assumed that the Galactic centre is located at a zenith angle $\zGC$ and azimuth $a=0^\circ$. To bin the neutrino flux, we consider 20 bins per decade as for the HKKM11 flux.  The number of energy bins we use depends on the DM mass, $E_\nu\in [10\MeV,\mdm]$ for DM annihilation into muons and $E_\nu\in [10\MeV,10^{\log(\lceil\mdm\rceil})]$ for annihilation into neutrinos, where $\lceil\mdm\rceil$ is the ceiling function. Note that in setting limits we combine the signal from all bins, resulting in an all-sky analysis.

\begin{table}[h]
    \centering
    \begin{tabular}{|l|c|c|}
    \hline
     & \multicolumn{2}{|c|}{$\mdm$ (GeV)} \\
    \hline
        Event category & $\chi\chi\rightarrow \mu^+\mu-$ &  $\chi\chi\rightarrow \nu \overline{\nu}$  \\
     \hline      
        FC $\nuebar$  & 0.11 - 50  &  0.017 - 50 \\  
        FC $\nue$  & 0.11 - 50  &  0.05 - 50 \\           
        FC $\numu$, $\numubar$ & 0.3 - 50  & 0.25 - 50 \\       
        PC $\numu$, $\numubar$ & 2 - 50 & 2 - 50 \\              
     \hline         
    \end{tabular}
    \caption{The dark matter mass ranges used for generating signal events in the different event categories. Each category also includes anti-neutrino events of the same flavour.}
    \label{tab:massrange}
\end{table}
The range of DM masses we use for the different event categories is shown in Table~\ref{tab:massrange}. We take an upper bound of 50~GeV in all cases in order to allow comparison with the HyperK Design Report. The lower bound varies from category to category. For muon final states the DM mass must be at least as large as $m_\mu$ from kinematic considerations. We find this absolute lower bound is only relevant for the FC $\nue + \nuebar$ category, and that the FC $\numu$ and PC $\numu$ classes do not have any acceptance below the DM masses shown in the table. For neutrino final states we choose the lower bound at 17~MeV, below which the experimental backgrounds are dominated by spallation. Again, this is only relevant in the FC $\nuebar$ class. As can be seen from Fig.~\ref{fig:EvsEkin} at low energies $\Ekin \lesssim 50$~MeV this event class is dominated by $\nuebar$ events, since the scattering cross-sections on hydrogen for $\nue$ are very suppressed. As expected, the partially contained category is associated with higher energy neutrinos (and hence DM masses).

\begin{figure}
\centering
\includegraphics[width=\textwidth]{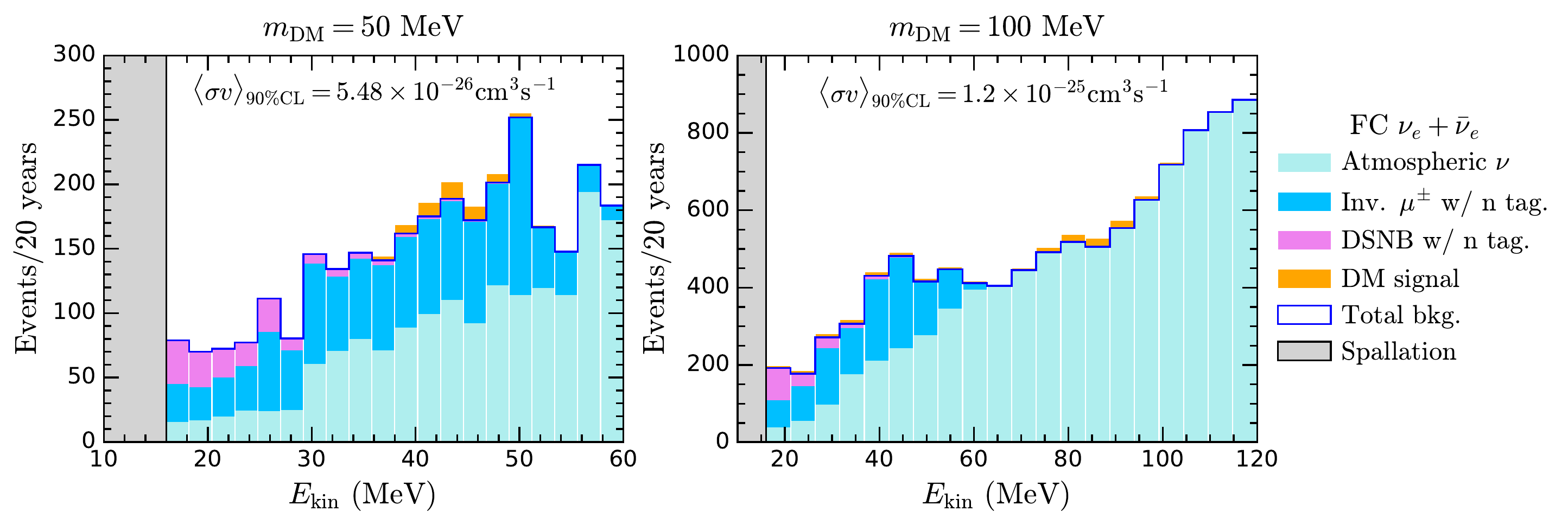}
\caption{Expected FC $\nue+\nuebar$ signal (orange) and background (blue) event rate at HyperK for DM annihilation into neutrinos, $\mdm=50\MeV$ (left) and $\mdm=100\MeV$ (right), and  20 years of livetime. The total background includes atmospheric neutrinos from HKKM11 and FLUKA (cyan), invisible muons (light blue) and DSNB (magenta). A 70\% neutron tagging efficiency is assumed. The DM signal corresponds to the 90\% CL $\sigmav$ upper bound shown in every panel.
Bins in the spallation region are not considered in the projected limit calculation. 
\label{fig:signalbkg}
}
\end{figure}

We use the \texttt{Swordfish} package~\cite{Edwards:2017mnf,Edwards:2017kqw} to derive 90\% confidence level (CL) upper bounds on the thermal annihilation cross-section. \texttt{Swordfish} calculates limits based on a maximum likelihood estimation, after reduction of the problem at hand to a single bin problem determined by an equivalent number of signal and background events.
For each DM mass and event class, we bin the all-sky data in $\Ekin$ between the lower threshold of 16~MeV and $1.2\,\mdm$. For masses above 40~MeV we use 20 bins and below 40~MeV we use 5 to avoid overbinning the data. We include an uncertainty of 10\% in the overall normalisation of the signal and backgrounds. This roughly agrees with the uncertainty in the atmospheric neutrino flux in~\cite{Abe:2017aap}.  We include the total background as discussed in the sections above, which includes the HKKM11 and FLUKA atmospheric neutrino fluxes, the invisible muon contribution and the DSNB. 

There is also considerable uncertainty around the exact normalisation of the DSNB. In our analysis we have simply fixed it to that used in the HyperK Design Report. However, a simultaneous analysis that takes into account the uncertainties due to the possible presence of dark matter and the uncertainties around the DSNB would be very interesting. The use of angular information could help, even though the angular resolution of HyperK is more limited at lower energies.

Fig.~\ref{fig:signalbkg} shows the size of the signal corresponding to the 90\% CL limit, together with the backgrounds, for two different choices of dark matter mass. Note that there are contributions to the signal at values of $\Ekin$ that are much lower than the DM mass.  This arises because, for the case of scattering from oxygen, $E_\nu$ and $\Ekin$ are not tightly correlated (see Fig.~\ref{fig:EvsEkin}).

\begin{figure}
    \centering
    \includegraphics[scale=0.75]{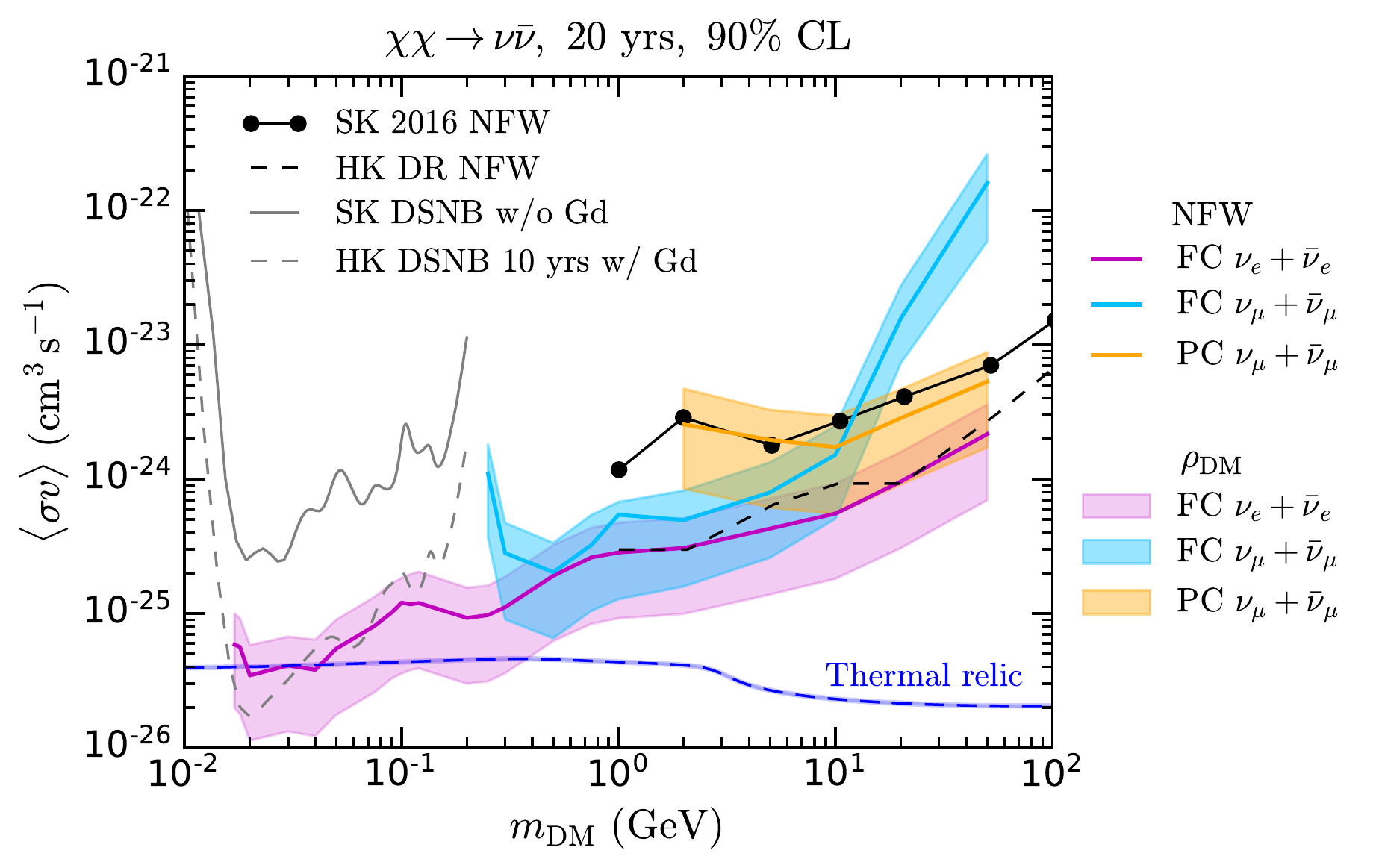}
    \caption{Limits derived for  $\chi\chi \to \nu\bar{\nu}$ annihilation for the event categories defined in the text for 20 years run time at HyperK. A 70\% neutron tagging efficiency is assumed. The central line in each band shows the limit for an NFW halo profile, with the upper and lower bounds being set by the limits for the Isothermal and Moore profiles respectively. We also show the limits from the SuperK GC dark matter search~\cite{Frankiewicz:2017trk} (solid with dots), and the projected limit from the HyperK DR~\cite{Abe:2018uyc} (dashed), a limit at low masses derived from the SuperK DSNB search~\cite{PalomaresRuiz:2007eu,Campo:2018dfh} (grey solid), projections for the low mass limit achievable in HyperK with Gd enhancement~\cite{Campo:2018dfh} (grey dashed)and the expected $\sigmav$ for a thermal relic calculated with \texttt{DarkSUSY} (blue dashed).\label{fig:nulimits}}
\end{figure}

\begin{figure}
    \centering
    \includegraphics[scale=0.75]{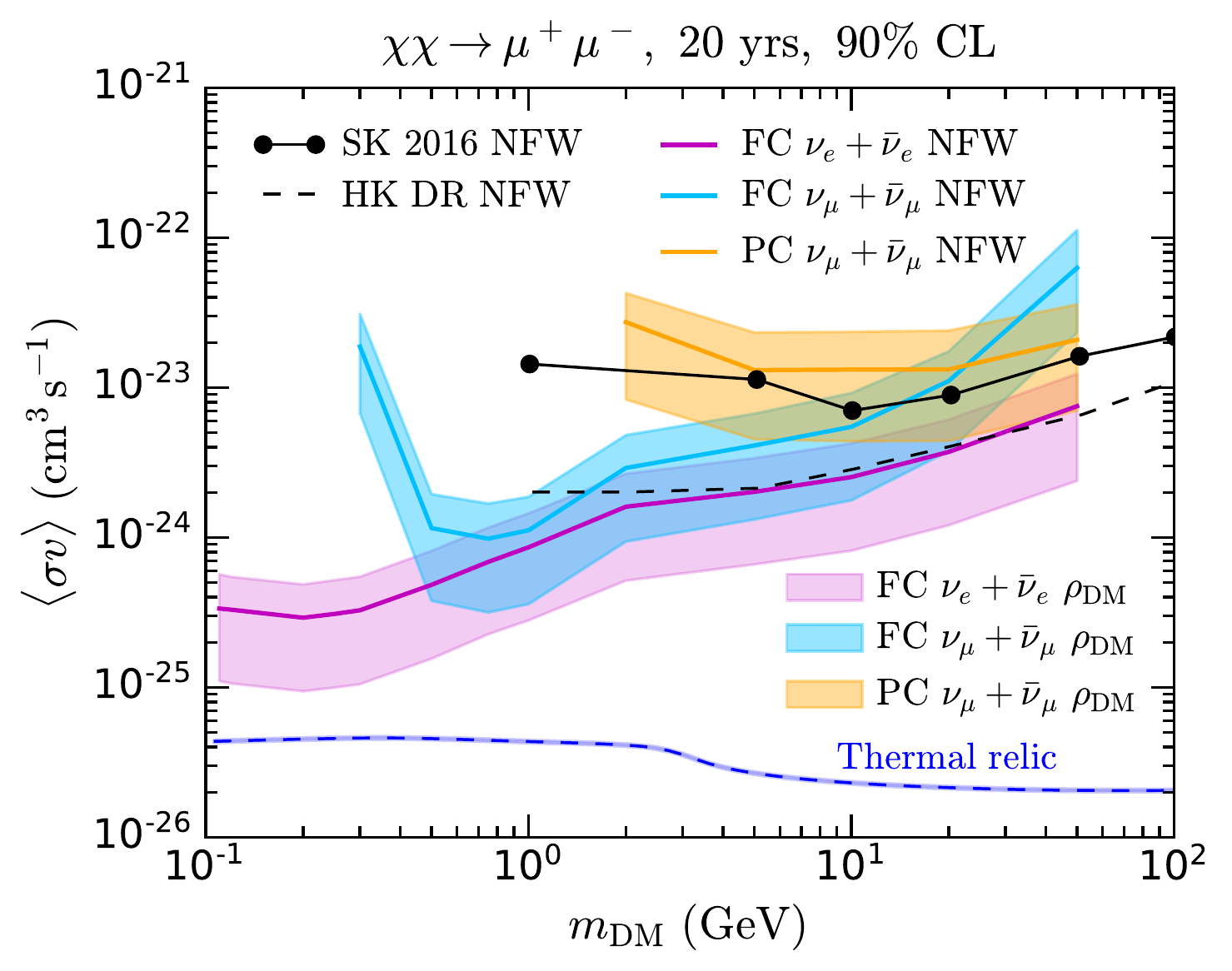}
    \caption{Limits derived for  $\chi\chi \to \mu^+ \mu^-$ annihilation for the event categories defined in the text for 20 years run time at HyperK. The central line in each band shows the limit for an NFW halo profile, with the upper and lower bounds being set by the limits for the Isothermal and Moore profiles respectively. We also show the limits from the SuperK GC dark matter search~\cite{Frankiewicz:2017trk} (solid with dots), and the projected limit from the HyperK Design Report~\cite{Abe:2018uyc} (dashed).\label{fig:mulimits}}
\end{figure}

We show our results for the neutrino and muon final states in Fig.~\ref{fig:nulimits} and~\ref{fig:mulimits} respectively. The solid black lines with points shows current bounds from the SuperK GC search~\cite{Frankiewicz:2017trk}
 and the dashed black line shows the projected limits from the HyperK DR~\cite{Abe:2018uyc}.  
 In Fig.~\ref{fig:nulimits} we also show a limit at low masses derived from the SuperK DSNB search~\cite{PalomaresRuiz:2007eu,Campo:2018dfh} as a grey solid line, and projections for the low mass limit achievable in HyperK with Gd enhancement~\cite{Campo:2018dfh} as a grey dashed line. Our limits for the FC $\nue+\nuebar$, FC $\numu+\numubar$ and PC $\numu+\numubar$ categories are shown as purple, light blue and orange regions respectively, with the FC $\nue+\nuebar$ category giving the strongest limit at low masses. The central line in each region corresponds to the limit for the NFW profile, while the upper and lower margins are the limits for  Isothermal and Moore  profiles respectively.  Note that both the Isothermal and Moore profiles examples are unrealistic choices for the halo profile. They serve as useful extremes to define a conservative error band that reflects uncertainty in the halo profile; the true uncertainty will be smaller.

 All our projections are for a 20 year running time, to facilitate comparison with the HyperK DR.
 In both plots we observe good agreement at high masses with the HyperK DR, indicating that our scaled-up version of the SuperK detector simulator captures sufficiently well the relevant features of HyperK. In Fig.~\ref{fig:nulimits} our results are broadly consistent with the projections of ref.~\cite{Campo:2018dfh}, which assume a 10 year running time.  This is also a non-trivial cross-check: their results are derived by re-interpreting the results of a SuperK DSNB search as a constraint on DM annihilation, and then rescaling those limits up to HyperK. 
 
 The thermal relic annihilation cross section \cite{Steigman:2012nb} is shown as the blue dashed line in Fig.~\ref{fig:nulimits}.  For the case of neutrino final states, we see that our projected sensitivity dips below the thermal annihilation cross-section of $\sim4\times 10^{-26}\cm^2$ at around 20~MeV, assuming the NFW halo profile. For muon final states the projected limits are approximately an order of magnitude higher than for neutrinos. Note, however, that annihilation to muons is subject to stronger constraints arising from the CMB at low mass and Fermi/AMS at high mass~\cite{Leane:2018kjk}.

\begin{figure}
    \centering
    \includegraphics[scale=0.75]{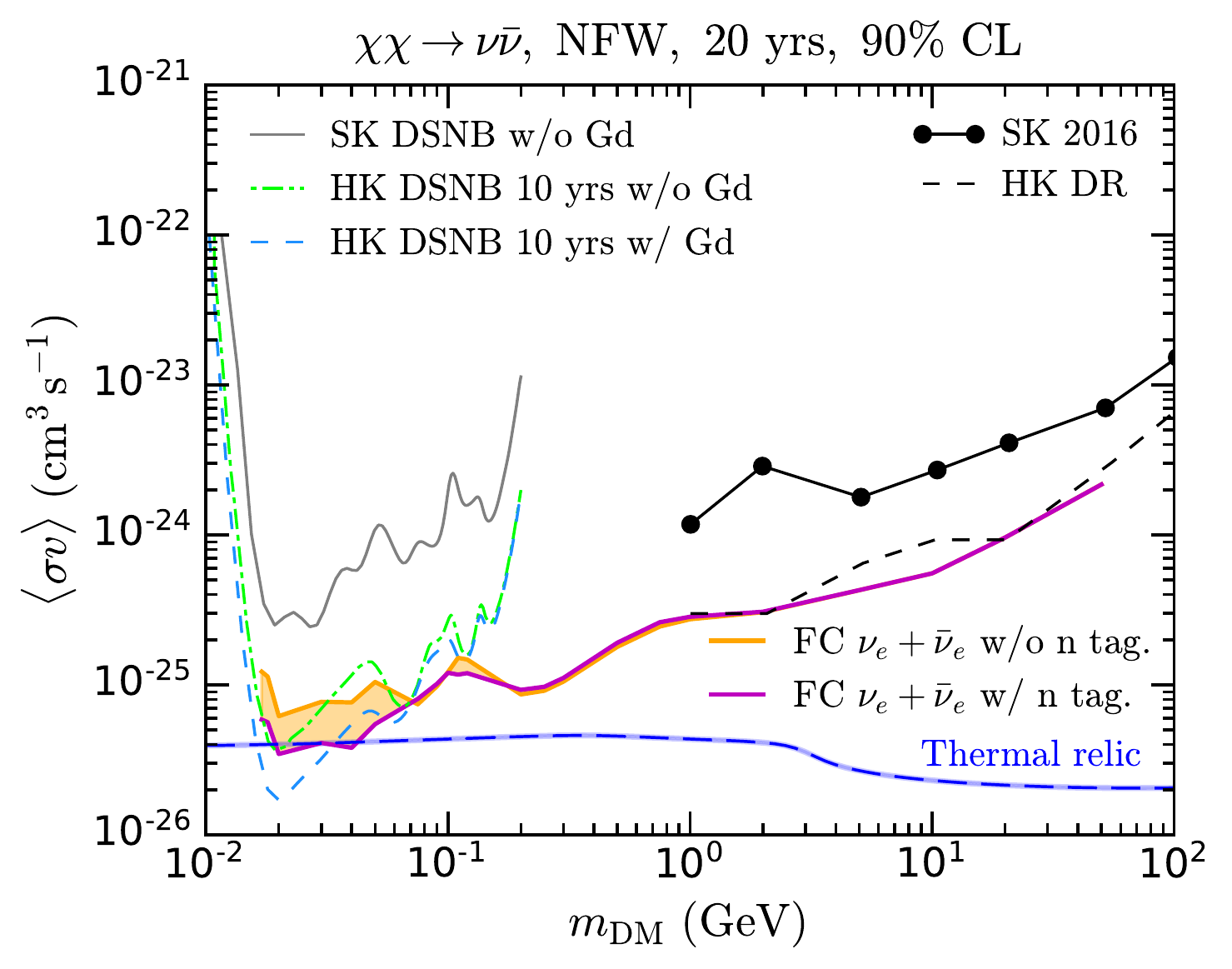}
    \caption{The limit on the annihilation cross-section, with and without neutron tagging, assuming a 70\% tagging efficiency. The purple line is identical to the limit from Fig.~\ref{fig:nulimits} for the NFW profile, while the orange line is the corresponding limit without neutron tagging. 
    The dashed light blue and dot-dashed light green lines show the projections from~\cite{Campo:2018dfh} for HyperK with and without neutron tagging respectively. The effects are confined to the low mass regime. \label{fig:limit_nfw_noGd}}
\end{figure}

Finally, in Fig.~\ref{fig:limit_nfw_noGd}, we compare the projected limits with and without neutron tagging, assuming annihilation into neutrinos and an NFW halo profile.  Since this mainly affects the invisible muon background at low energies, we only show the FC $\nue+\nuebar$ event category in this plot. Neutron tagging would have a negligible effect on the limits for annihilation into muons. The impact on the projected sensitivity for annihilation to neutrinos is seen below 70~MeV, where the limit without n-tagging (shown as an orange line) becomes weaker by a factor of about two relative to the n-tagged case. We also show the projected limits with (dashed light blue) and without (dot-dashed light green) n-tagging at HyperK from ref.~\cite{Campo:2018dfh}.

\section{Conclusions}
\label{sec:conc}

We have projected the reach of the Hyper-Kamiokande experiment in searches for light dark matter that annihilates to neutrinos and muons. To generate our results we have used an original detector simulation validated against results from Super-Kamiokande and then scaled up according to parameters in the HyperK Design Report~\cite{Abe:2018uyc}. We have focused on an annihilation signal originating in the centre of the Galaxy (although we technically undertake an all-sky analysis neglecting the extragalactic signal contribution) finding that HyperK should be able to probe thermal annihilation cross-sections for DM masses around 20~MeV for annihilations into neutrinos (for an NFW halo). The sensitivity for muon final states is always at least an order of magnitude above the thermal cross-section. We note that there are substantial uncertainties in these projections depending on the exact form of the DM halo profile, and possible galactic substructure.

At low masses a critical background derives from muon-induced spallation products. Based on information in the HyperK DR~\cite{Abe:2018uyc}, we have adopted 17~MeV as the lower threshold for our projections. However, neutron tagging would allow these spallation events to be identified~\cite{Li:2014sea,Li:2015kpa,Li:2015lxa}, possibly allowing a search down to neutrino energies of 10~MeV. While the current limits on thermal dark matter annihilating into neutrinos from the Cosmic Microwave Background and Big Bang Nucleosynthesis are $\mathcal{O}(1-10)$~MeV\footnote{The precise number depends on the dark matter spin and other properties.}, CMB Stage-IV experiments should be able to probe masses up to 10-15~MeV~\cite{Escudero:2018mvt,Sabti:2019mhn}. The upcoming JUNO and DUNE experiments are also projected to have sensitivity to thermal cross-sections at low masses~\cite{Klop:2018ltd,Arguelles:2019ouk}. The combination of HyperK, CMB Stage-IV, JUNO and DUNE data may thus allow a comprehensive probe of dark matter annihilating to neutrinos up to masses of several tens of MeV, substantially extending the reach of current experiments.

We thus consider it important that HyperK undertakes searches for dark matter over the full range of experimental sensitivity, and in particular down to the lowest energies. The prospects for DM discovery at HyperK would be further enhanced with the presence of a second detector in South Korea~\cite{Abe:2016ero}. Although the main advantages of such a detector would be in neutrino oscillation physics, the limits on the dark matter annihilation cross-section could be expected to improve by an $\mathcal{O}(1)$ factor, assuming identical detectors and exposure times. We note that the larger overburdens at the proposed South Korean sites would be particularly beneficial in searches for light dark matter through reducing the spallation backgrounds at low energies. 

This paper opens a number of possible directions for future work through improving our simulations and extending the signatures studied.  The Hyper-Kamiokande experiment will be an exciting new tool in the quest to understand dark matter, and we plan to return to these issues in the future.

\acknowledgments
 We thank John Beacom for detailed comments on the manuscript and Jost Migenda and Tommaso Boschi for helpful discussions and comments. NFB, MJD and SR are supported by the Australian Research Council. This research was supported by the Munich Institute for Astro and Particle Physics (MIAPP) which is funded by the  Deutsche  Forschungsgemeinschaft  (DFG)   under   Germany’s   Excellence Strategy  EXC-2094  390783311.

\appendix
\section{Bonus Plots}
\label{sec:app}
\begin{figure}
    \centering
     \includegraphics[scale=0.49]{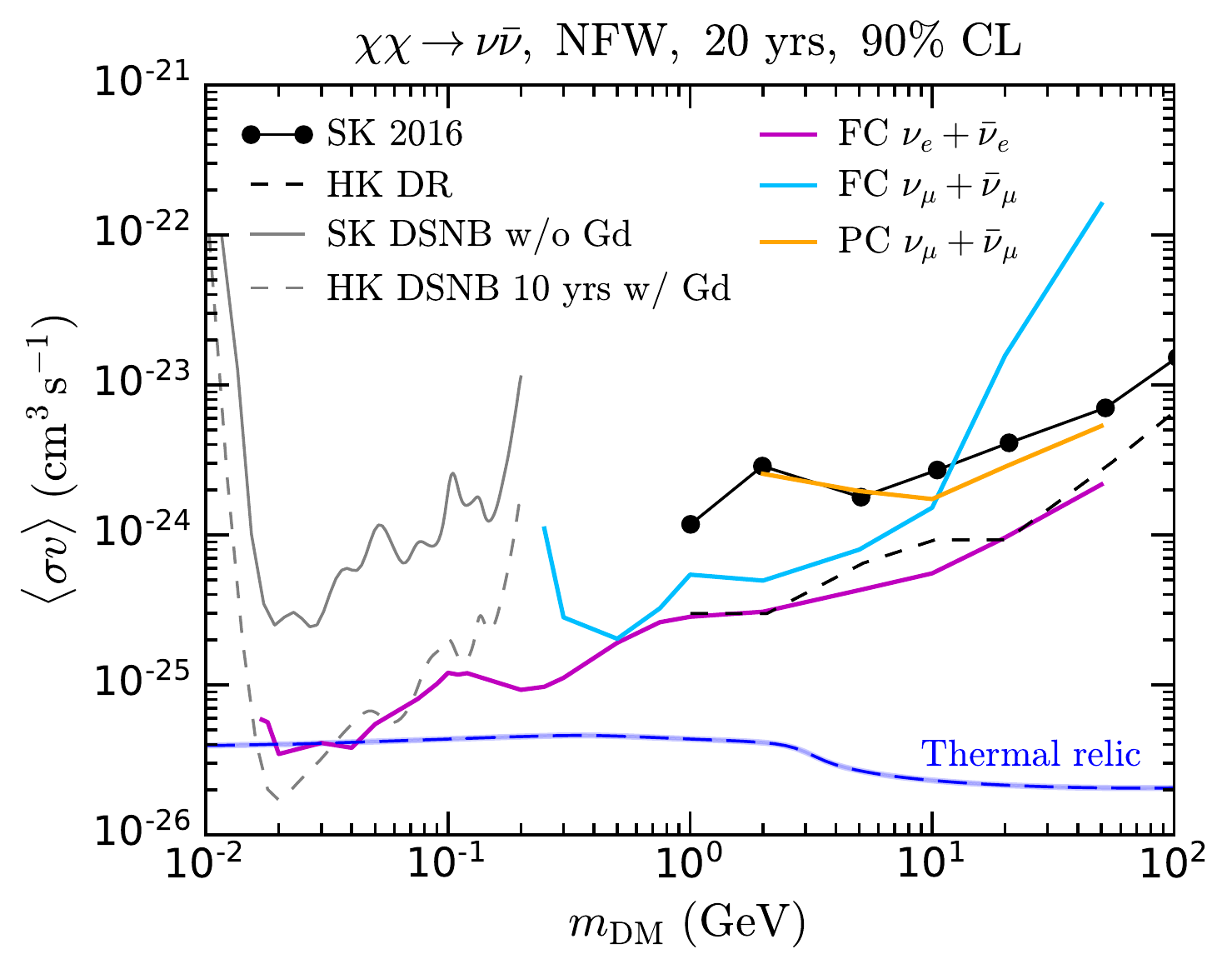}
    \includegraphics[scale=0.49]{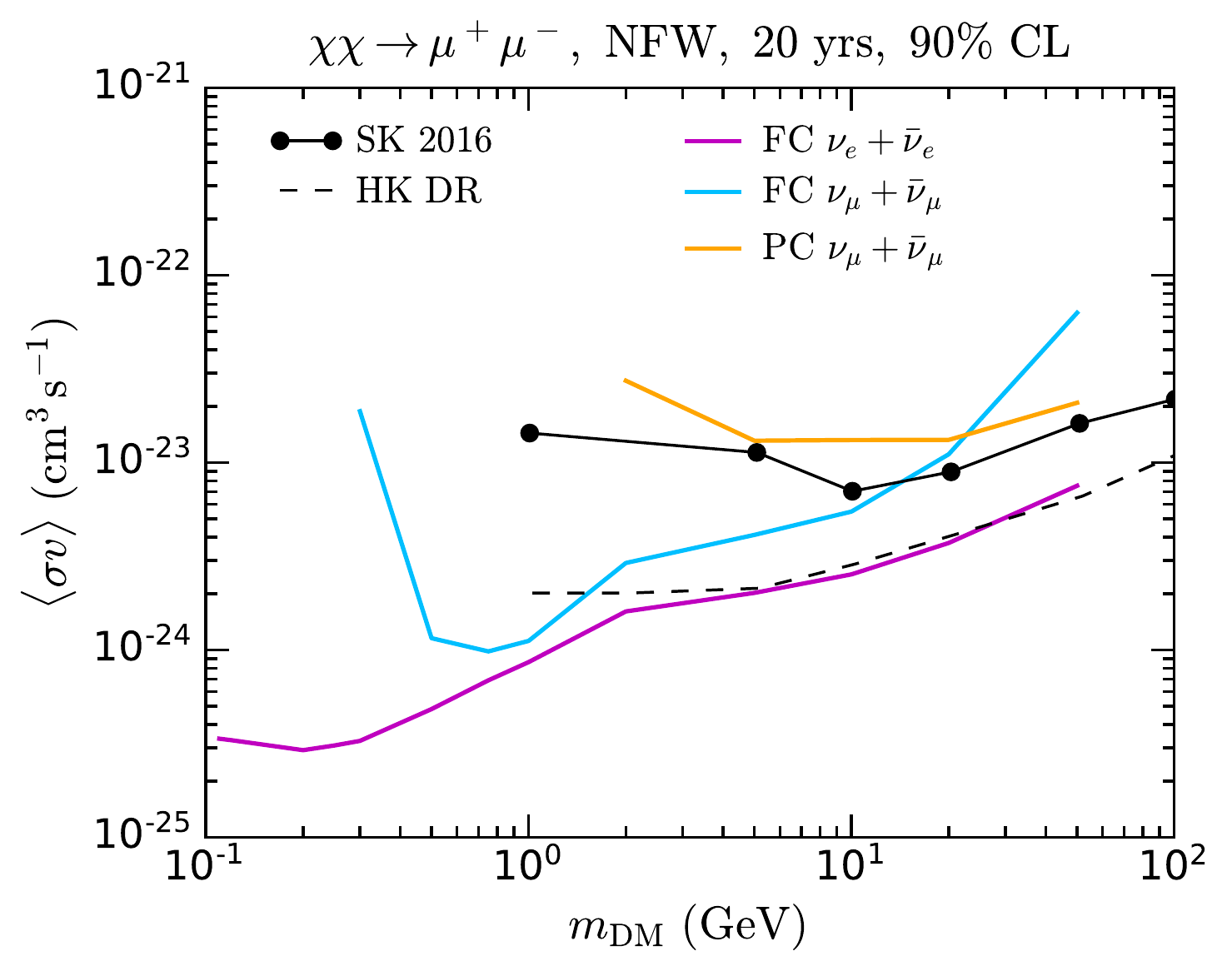}
    \includegraphics[scale=0.49]{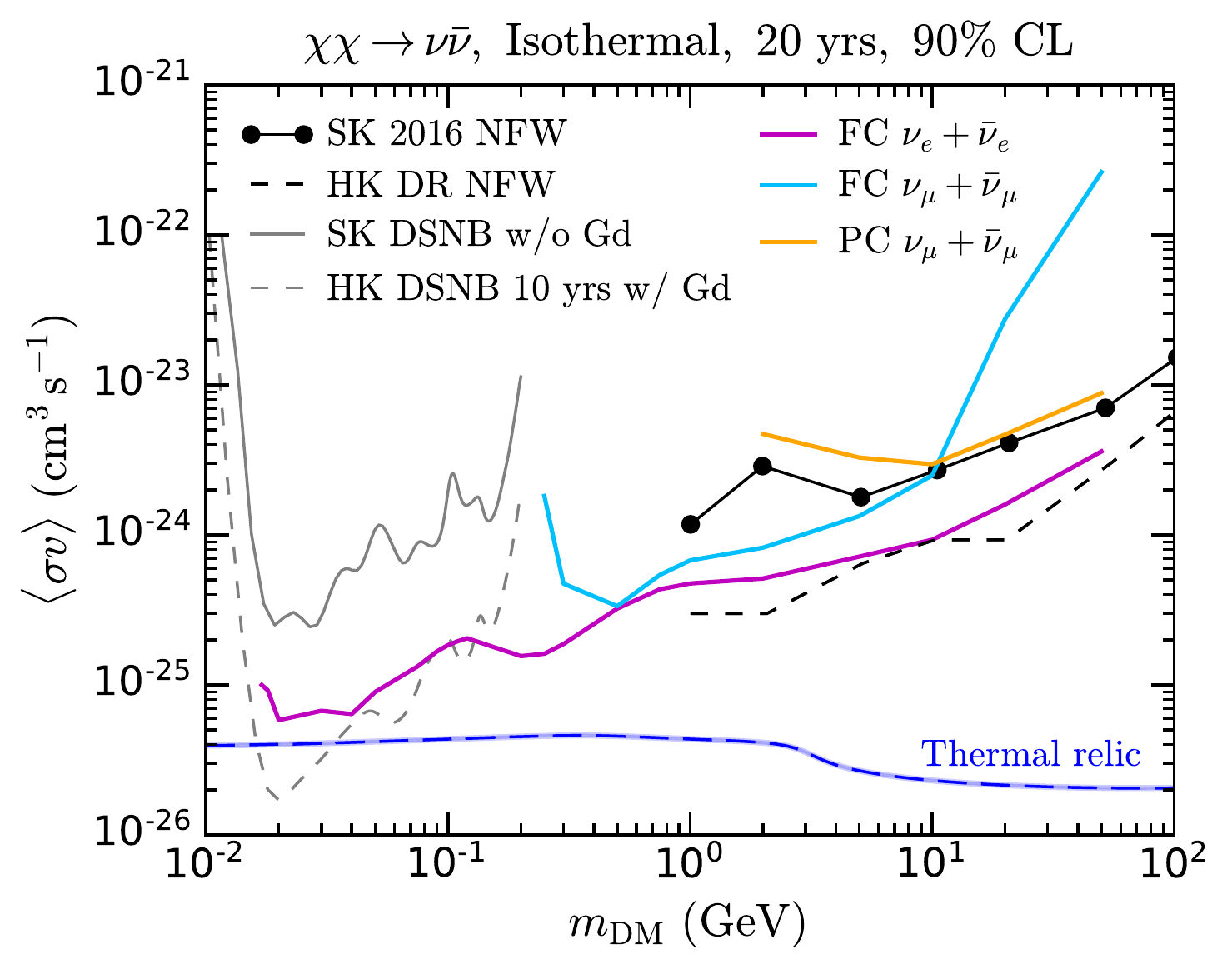}
    \includegraphics[scale=0.49]{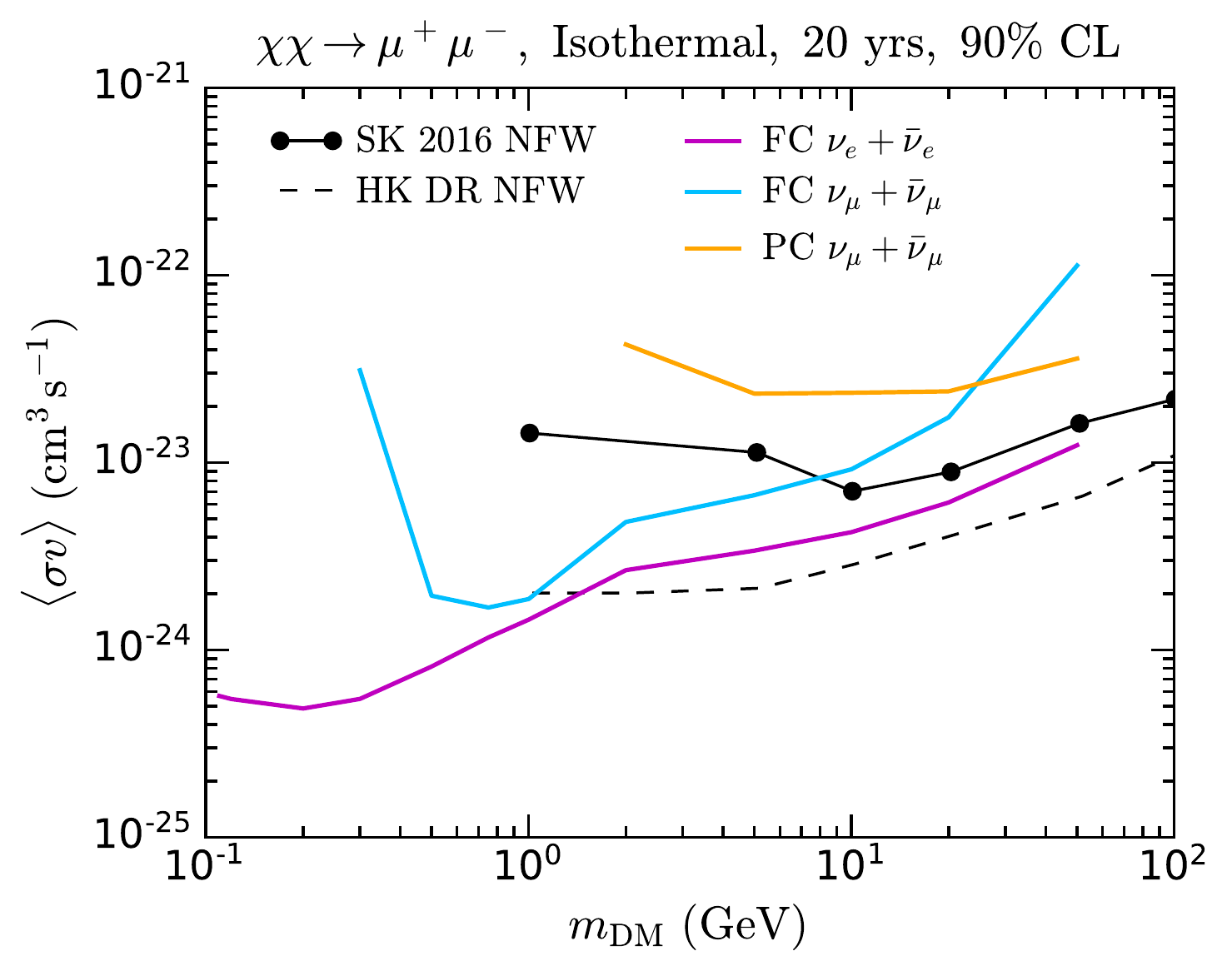}
     \includegraphics[scale=0.49]{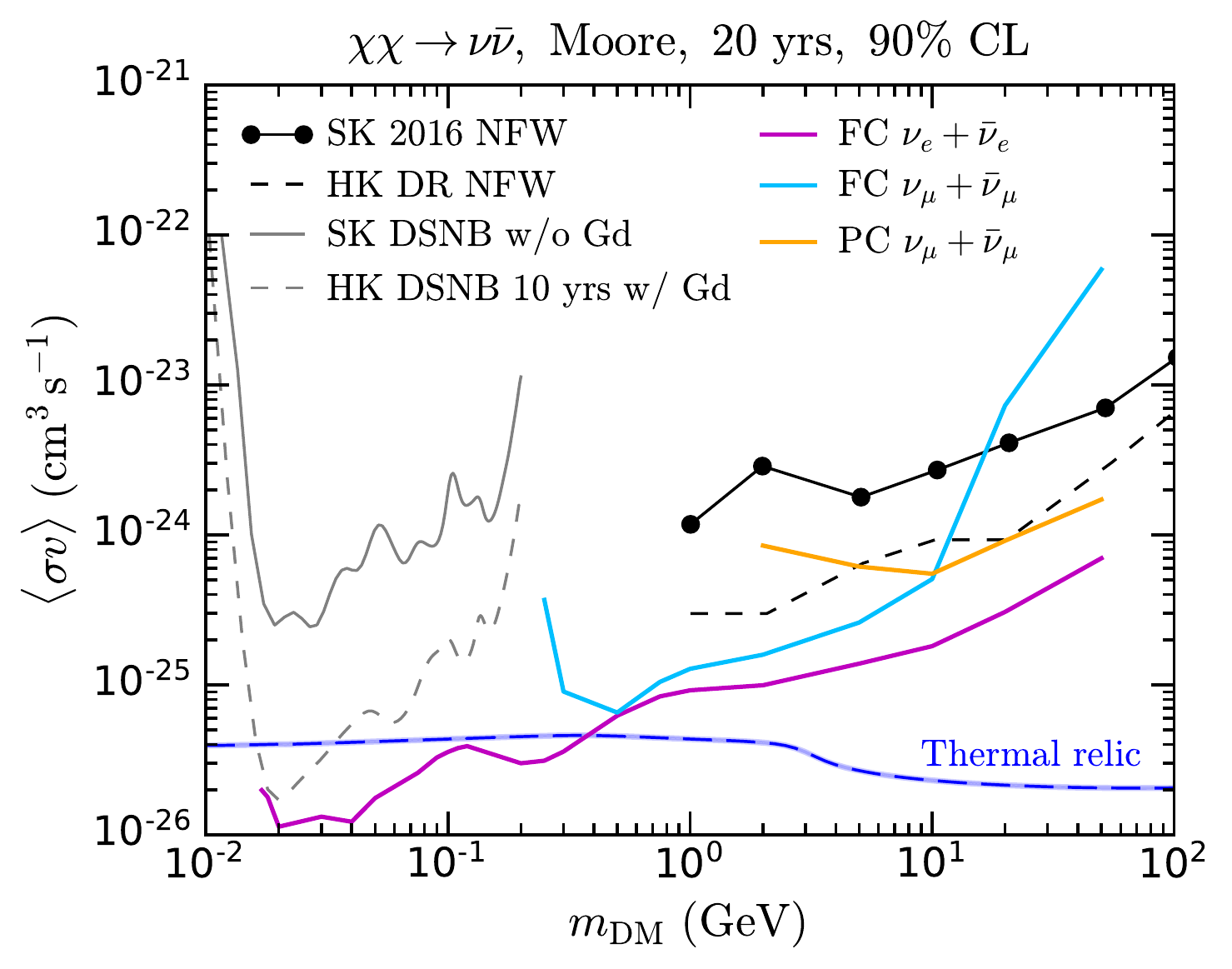}
    \includegraphics[scale=0.49]{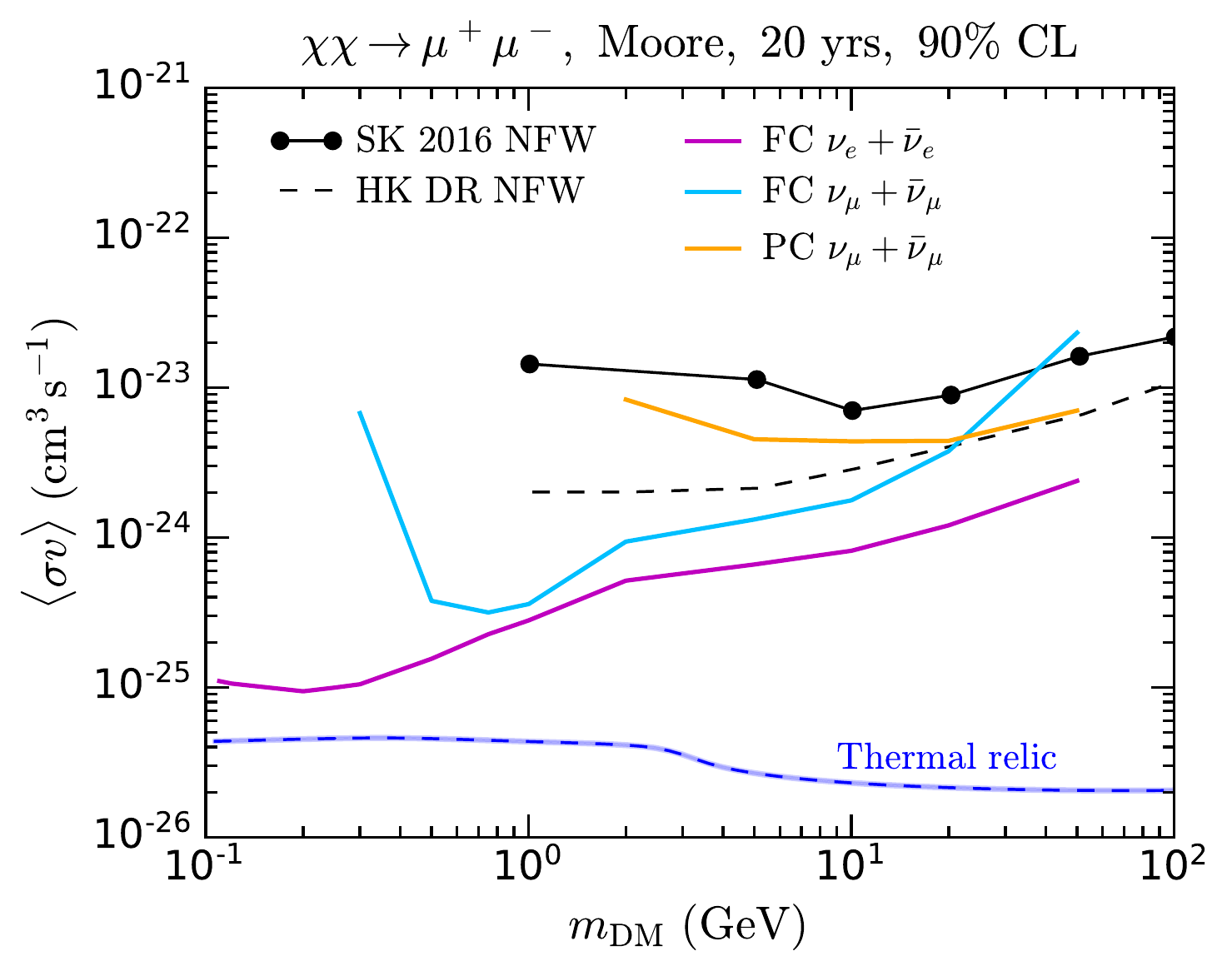}
    \caption{Limits plots for the $\mu^+\mu^-$ (left) and $\nu\overline{\nu}$ (right) final states for all considered dark matter halo profiles, namely NFW (top row),  Isothermal (middle row) and Moore (bottom row). Halo profiles are defined in Section~\ref{sec:limits}.}
    \label{fig:indiv_limits}
\end{figure}

\begin{figure}
    \centering
    \includegraphics[scale=0.49]{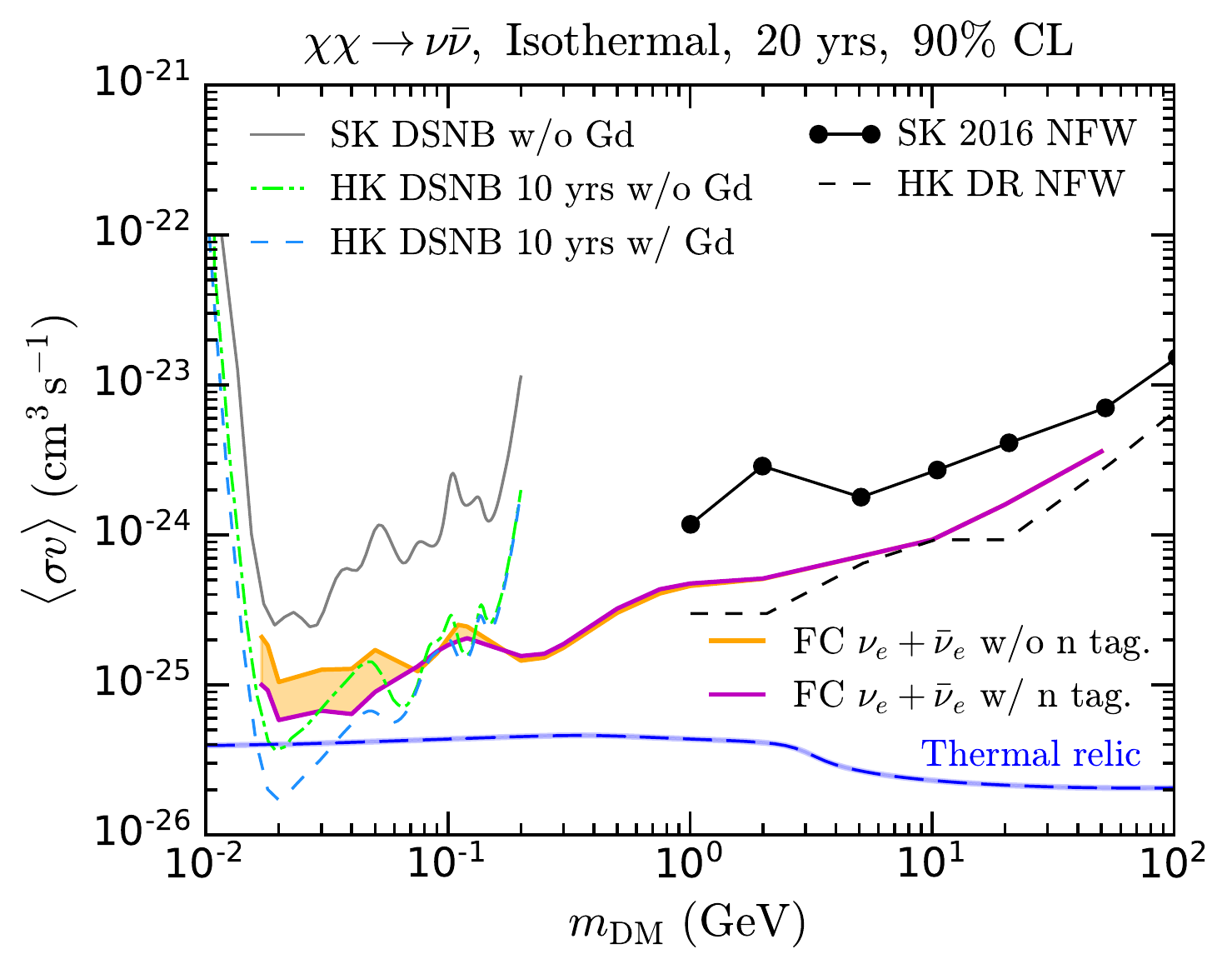}    
    \includegraphics[scale=0.49]{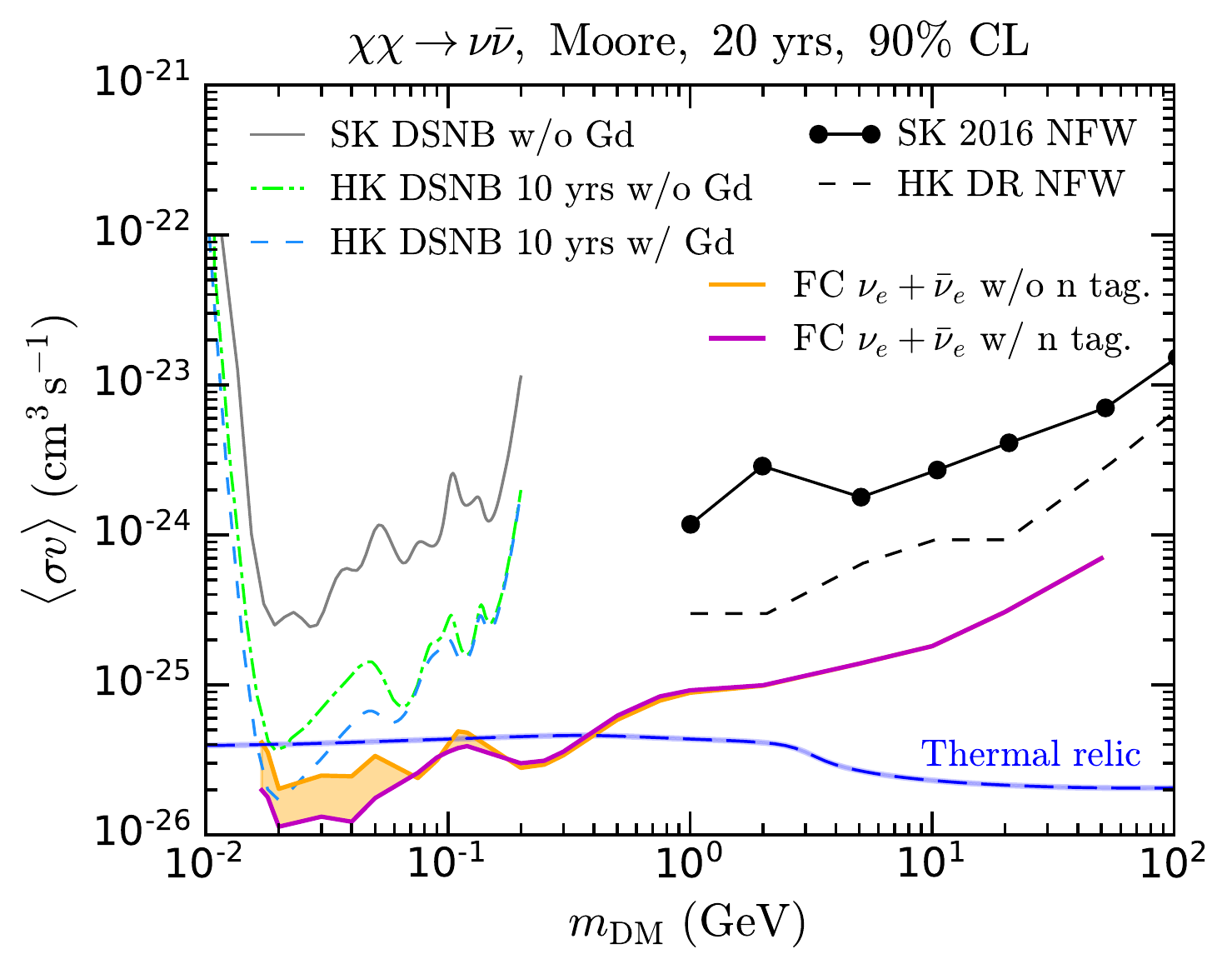}       
    \caption{Projected limits for the Isothermal (left) and Moore (right) halos, with and without neutrino tagging, assuming a 70\% tagging efficiency. These plots assume the same 20 year exposure time as our other results.\label{fig:limits_althalo_noGd}}
\end{figure}

For completeness and clarity, in this section we show individually in Fig~\ref{fig:indiv_limits} the limits we obtain for the NFW (top row), Isothermal (middle row) and Moore profiles (bottom row). 
In Fig.~\ref{fig:limits_althalo_noGd} we also show our projections for the Isothermal (left panel) and Moore (right panel) halos, with and without including the effects of neutrino-tagging in the HyperK detector.

\section{Time-averaged J-factor and matter oscillation effects}
\label{sec:app2}

For simplicity in our main results above we placed the GC at the horizon. Since the atmospheric neutrino background is also correlated with the horizon one might wonder if this has a large impact on our results. In this appendix we demonstrate that this is not the case. We average the position of the GC over a twenty-four hours period, and show that this does not demonstrably affect our signal. 

To properly calculate the J-factor in horizontal coordinates at the HyperK site, we have first performed the transformation from Galactic  $(l,b)$ to equatorial coordinates $(\rm RA^\circ,DEC^\circ)$, using the 
equatorial coordinates of the North Galactic Pole  for the J2000 equinox  from ref.~\cite{Reid:2004rd}, the  same as in {\tt Astropy}~\cite{astropy:2013,astropy:2018} and the North Celestial Pole longitude  from {\tt Astropy}. Next, we have converted $(\rm RA^\circ,DEC^\circ)$ to $(a,z)$, using the HyperK geographic coordinates in Table~\ref{tab:detectorparameters} of Ref.~\cite{Abe:2018uyc} and  the Greenwich apparent sidereal time (GAST) (including both Earth procession and nutation)  computed with the Standard of Fundamental Astronomy  {\tt SOFA} C library \cite{SOFA:2019-07-22} from the International Astronomical Union. 
 To obtain the GAST we require the time correction $\rm DUT_1=UT1-UTC$  for a specific day in leap seconds. We have used the prediction for June 1st, 2021 from the International Earth Rotation and Reference Systems Service Bulletin~A\footnote{\url{https://datacenter.iers.org/data/latestVersion/6_BULLETIN_A_V2013_016.txt}}.
 
\begin{figure}[h]
    \centering
\includegraphics[width=0.49\textwidth]{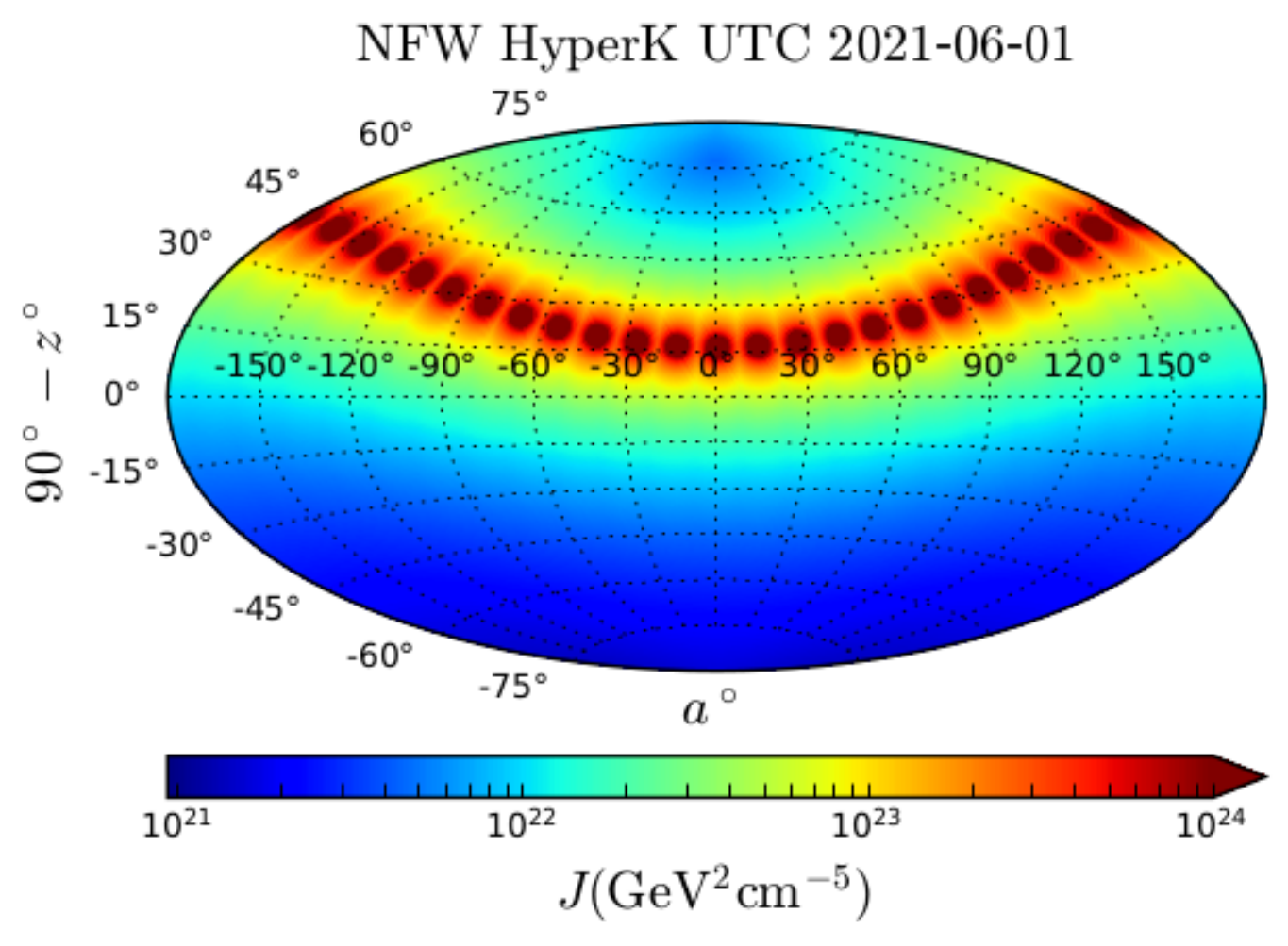}
\includegraphics[width=0.49\textwidth]{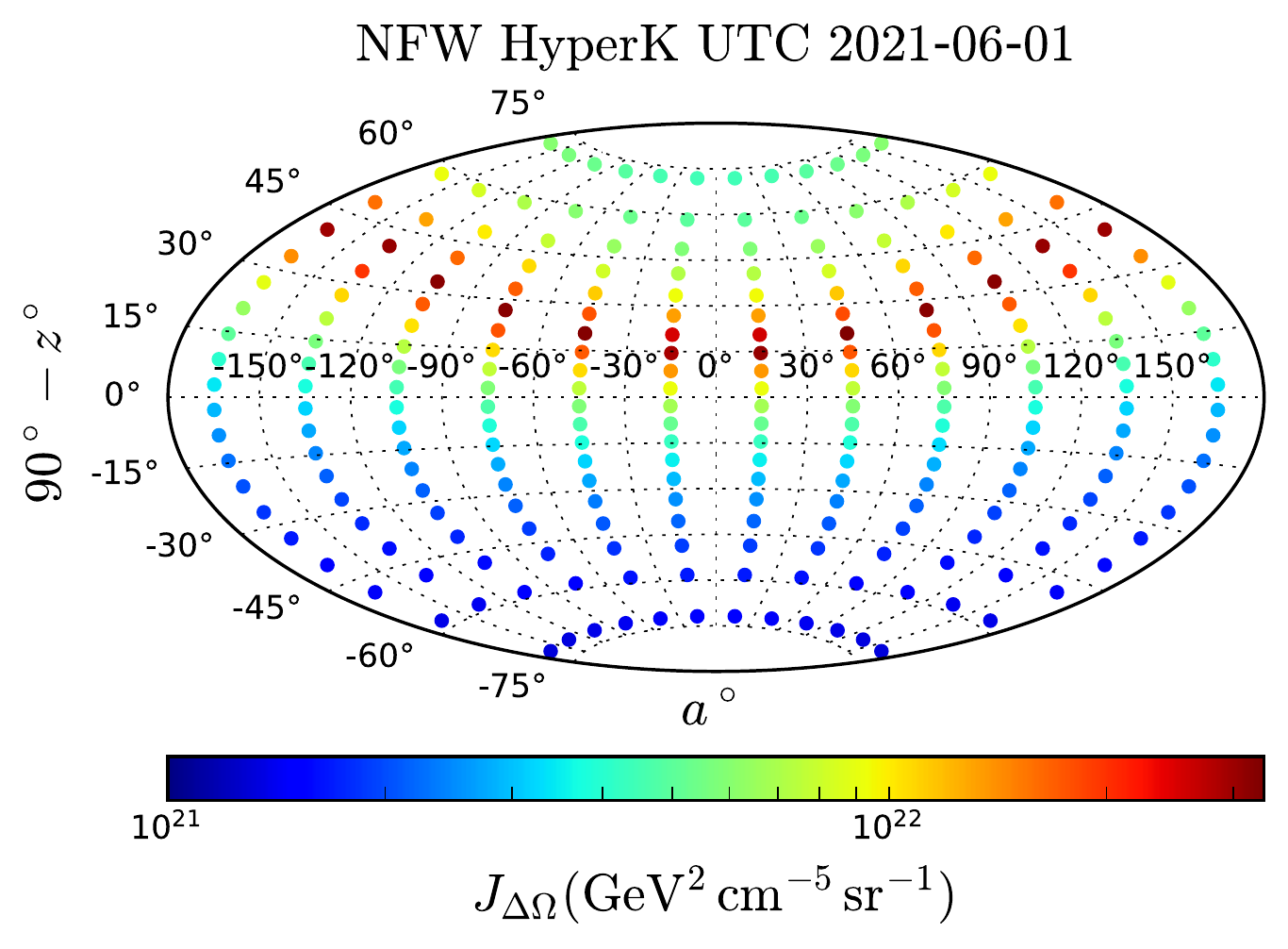}
    \caption{Left: J-factor in horizontal coordinates at HyperK site, prediction per hour for June 1st, 2021. Right: Same J factor binned in azimuth and zenith angles and averaged over 24 hours.}
    \label{fig:dayaveJfactor}
\end{figure}

 We have computed $J(a,z)$ for every hour of the aforementioned day, averaged it over 24 hours and binned it using Eq.~\ref{eq:Jfactorbinned}. The left panel of Fig.~\ref{fig:dayaveJfactor} shows the J-factor calculated in this way. The right panel of Fig.~\ref{fig:dayaveJfactor} shows the same J-factor binned using Eq.~\ref{eq:Jfactorbinned}.

\begin{figure}[h]
\centering
\includegraphics[width=\textwidth]{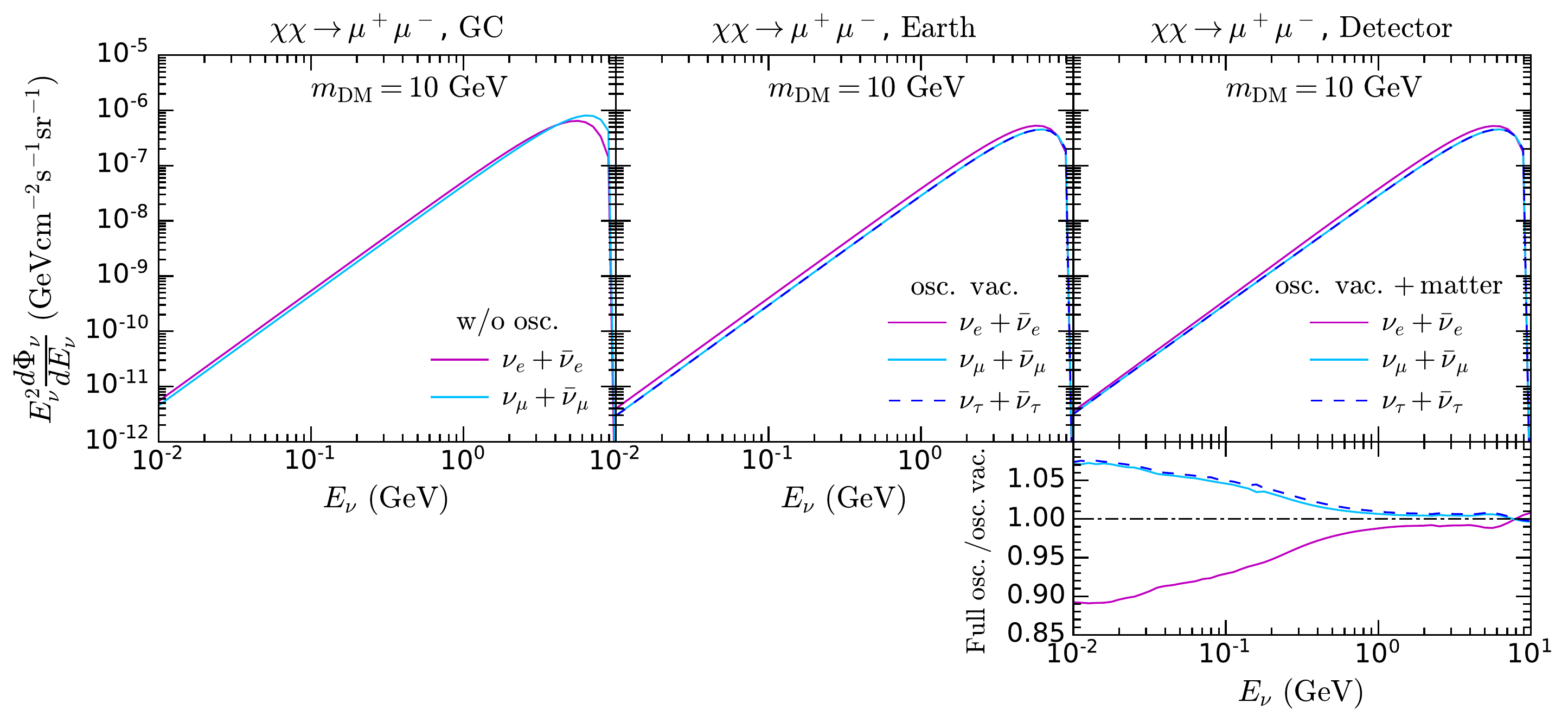}
\caption{
Galactic DM induced neutrino flux in the GC (left), at the Earth (with oscillations in vacuum, middle) and the detector depth (with oscillations in vacuum and in matter, right), for a generic WIMP with NFW DM profile $\mdm=10\GeV$  and  $\sigmav=2.31\times 10^{-26}\cm^3\s^{-1}$ . The bottom right panel shows the ratio between the neutrino flux at the detector depth (with oscillation in vacuum and matter) and the flux at Earth, which accounts only for oscillations in vacuum. }
\label{fig:matter}
\end{figure}

We have also considered the impact of matter oscillations on our signal. The left panel of Fig.~\ref{fig:matter} shows the spectrum of neutrinos at the GC (after the muons have decayed) obtained with Eq.~\ref{eq:GCneuflux}, for $\mdm=10$~GeV. The middle plot shows the neutrino spectrum at the Earth, including vacuum oscillations, Eq.~\ref{eq:DMneuoscvac}. The third plot shows the spectrum at the HyperK site, including matter effects, calculated with \texttt{nuCraft} as outlined in section~\ref{sec:geometry}. The fourth, bottom right panel shows the ratio of each flavour of flux at detector depth to the corresponding flux at Earth. We see that matter oscillations are negligible around the peak of the distribution, and are a 5-10\% percent effect in the low energy tail. We therefore do not expect matter effects to have a substantial impact on our projections.

\begin{figure}[h]
    \centering
\includegraphics[width=0.49\textwidth]{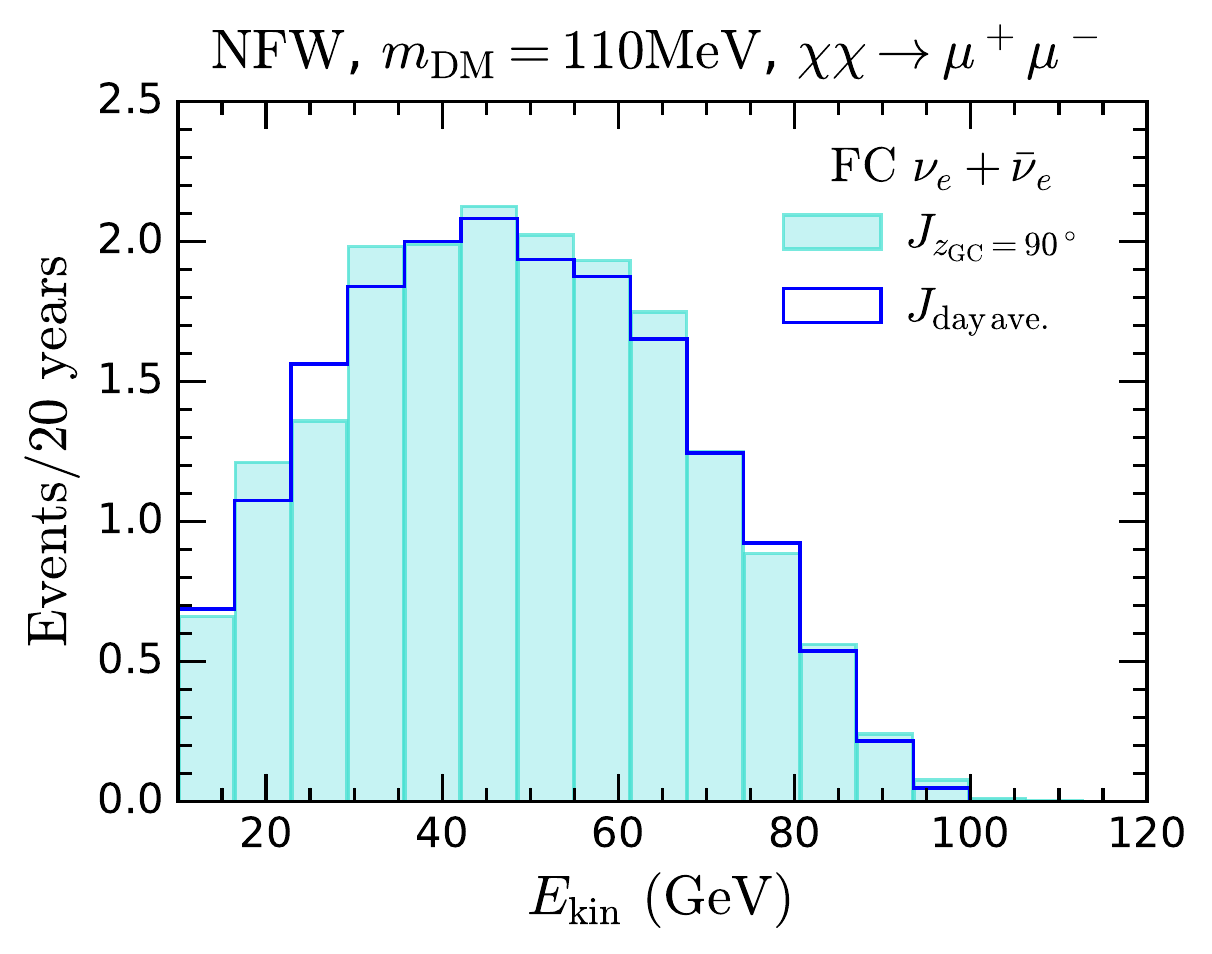}
\includegraphics[width=0.49\textwidth]{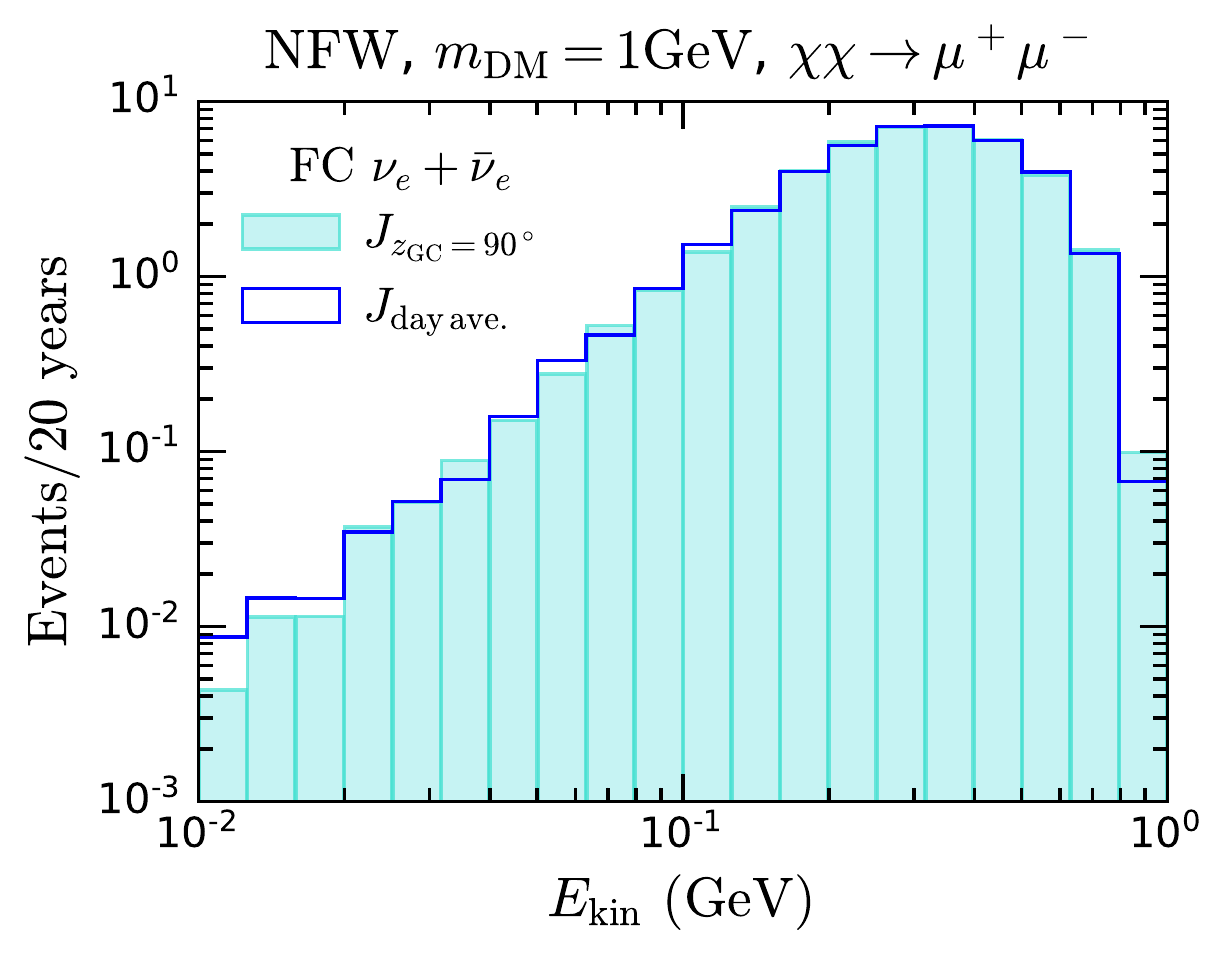}
    \caption{Expected FC $\nue+\nuebar$ signal event rate at HyperK for DM annihilation into muons, for  $\mdm=110\MeV$ (left) and  $\mdm=1\GeV$ (right), 
with $\sigmav=4.38\times 10^{-26}\cm^3\s^{-1}$ and $\sigmav=4.36\times 10^{-26}\cm^3\s^{-1}$ respectively, and 20 years of livetime. The event rate calculated using the DM-induced neutrino flux with the J factor binned in horizontal coordinates, the GC at the horizon $\zGC=90^\circ$ (Fig.~\ref{fig:Jfactorbinned}) and considering only oscillations in vacuum is shown in cyan. Blue lines correspond to event rates obtained with fluxes computed with the 
    binned day-averaged $J$ factor in Fig.~\ref{fig:dayaveJfactor} and including oscillations in vacuum and in matter. }
   \label{fig:sig_newJfactor}
\end{figure}

In Fig.~\ref{fig:sig_newJfactor} we show the expected FC $\nue+\nuebar$ signal event rate at HyperK for DM annihilation into muons, for  $\mdm=110\MeV$ (left) and  $\mdm=1\GeV$ (right). The plots are normalised to $\sigmav=4.38\times 10^{-26}\cm^3\s^{-1}$ and $\sigmav=4.36\times 10^{-26}\cm^3\s^{-1}$ respectively, corresponding to the required annihilation cross-section to achieve the observed relic density, and 20 years of live time. The event rate calculated assuming the GC at the horizon $\zGC=90^\circ$, as we use throughout the main text, is shown as filled histogram in cyan. The cyan bins also neglect the impact of matter oscillations.

The solid blue lines correspond to event rates obtained with fluxes computed with the 
    binned day-averaged $J$ factor in Fig.~\ref{fig:dayaveJfactor} and including oscillations in vacuum and in matter. The differences between the plots are minor, justifying our use of the GC at the horizon in the main text.

\bibliographystyle{JHEP} 
\bibliography{references}

\providecommand{\href}[2]{#2}\begingroup\raggedright\begin{thebibliography}{10}

\bibitem{Leane:2018kjk}
R.~K. Leane, T.~R. Slatyer, J.~F. Beacom and K.~C.~Y. Ng, \emph{{GeV-scale
  thermal WIMPs: Not even slightly ruled out}},
  \href{http://dx.doi.org/10.1103/PhysRevD.98.023016}{\emph{Phys. Rev.} {\bf
  D98} (2018) 023016}, [\href{http://arxiv.org/abs/1805.10305}{{\tt
  1805.10305}}].

\bibitem{Beacom:2006tt}
J.~F. Beacom, N.~F. Bell and G.~D. Mack, \emph{{General Upper Bound on the Dark
  Matter Total Annihilation Cross Section}},
  \href{http://dx.doi.org/10.1103/PhysRevLett.99.231301}{\emph{Phys. Rev.
  Lett.} {\bf 99} (2007) 231301},
  [\href{http://arxiv.org/abs/astro-ph/0608090}{{\tt astro-ph/0608090}}].

\bibitem{Yuksel:2007ac}
H.~Yuksel, S.~Horiuchi, J.~F. Beacom and S.~Ando, \emph{{Neutrino Constraints
  on the Dark Matter Total Annihilation Cross Section}},
  \href{http://dx.doi.org/10.1103/PhysRevD.76.123506}{\emph{Phys. Rev.} {\bf
  D76} (2007) 123506}, [\href{http://arxiv.org/abs/0707.0196}{{\tt
  0707.0196}}].

\bibitem{Boehm:2013jpa}
C.~Boehm, M.~J. Dolan and C.~McCabe, \emph{{A Lower Bound on the Mass of Cold
  Thermal Dark Matter from Planck}},
  \href{http://dx.doi.org/10.1088/1475-7516/2013/08/041}{\emph{JCAP} {\bf 1308}
  (2013) 041}, [\href{http://arxiv.org/abs/1303.6270}{{\tt 1303.6270}}].

\bibitem{Nollett:2013pwa}
K.~M. Nollett and G.~Steigman, \emph{{BBN And The CMB Constrain Light,
  Electromagnetically Coupled WIMPs}},
  \href{http://dx.doi.org/10.1103/PhysRevD.89.083508}{\emph{Phys. Rev.} {\bf
  D89} (2014) 083508}, [\href{http://arxiv.org/abs/1312.5725}{{\tt
  1312.5725}}].

\bibitem{Nollett:2014lwa}
K.~M. Nollett and G.~Steigman, \emph{{BBN And The CMB Constrain Neutrino
  Coupled Light WIMPs}},
  \href{http://dx.doi.org/10.1103/PhysRevD.91.083505}{\emph{Phys. Rev.} {\bf
  D91} (2015) 083505}, [\href{http://arxiv.org/abs/1411.6005}{{\tt
  1411.6005}}].

\bibitem{Escudero:2018mvt}
M.~Escudero, \emph{{Neutrino decoupling beyond the Standard Model: CMB
  constraints on the Dark Matter mass with a fast and precise $N_{\rm eff}$
  evaluation}},
  \href{http://dx.doi.org/10.1088/1475-7516/2019/02/007}{\emph{JCAP} {\bf 1902}
  (2019) 007}, [\href{http://arxiv.org/abs/1812.05605}{{\tt 1812.05605}}].

\bibitem{Sabti:2019mhn}
N.~Sabti, J.~Alvey, M.~Escudero, M.~Fairbairn and D.~Blas, \emph{{Refined
  Bounds on MeV-scale Thermal Dark Sectors from BBN and the CMB}},
  \href{http://dx.doi.org/10.1088/1475-7516/2020/01/004}{\emph{JCAP} {\bf 2001}
  (2020) 004}, [\href{http://arxiv.org/abs/1910.01649}{{\tt 1910.01649}}].

\bibitem{Campo:2017nwh}
A.~Olivares-Del~Campo, C.~Bœhm, S.~Palomares-Ruiz and S.~Pascoli, \emph{{Dark
  matter-neutrino interactions through the lens of their cosmological
  implications}},
  \href{http://dx.doi.org/10.1103/PhysRevD.97.075039}{\emph{Phys. Rev.} {\bf
  D97} (2018) 075039}, [\href{http://arxiv.org/abs/1711.05283}{{\tt
  1711.05283}}].

\bibitem{Elor:2018twp}
G.~Elor, M.~Escudero and A.~Nelson, \emph{{Baryogenesis and Dark Matter from
  $B$ Mesons}}, \href{http://dx.doi.org/10.1103/PhysRevD.99.035031}{\emph{Phys.
  Rev. D} {\bf 99} (2019) 035031}, [\href{http://arxiv.org/abs/1810.00880}{{\tt
  1810.00880}}].

\bibitem{Blennow:2019fhy}
M.~Blennow, E.~Fernandez-Martinez, A.~Olivares-Del~Campo, S.~Pascoli,
  S.~Rosauro-Alcaraz and A.~V. Titov, \emph{{Neutrino Portals to Dark Matter}},
  \href{http://dx.doi.org/10.1140/epjc/s10052-019-7060-5}{\emph{Eur. Phys. J.}
  {\bf C79} (2019) 555}, [\href{http://arxiv.org/abs/1903.00006}{{\tt
  1903.00006}}].

\bibitem{Ballett:2019pyw}
P.~Ballett, M.~Hostert and S.~Pascoli, \emph{{Dark Neutrinos and a Three Portal
  Connection to the Standard Model}},
  \href{http://arxiv.org/abs/1903.07589}{{\tt 1903.07589}}.

\bibitem{Boehm:2000gq}
C.~Boehm, P.~Fayet and R.~Schaeffer, \emph{{Constraining dark matter candidates
  from structure formation}},
  \href{http://dx.doi.org/10.1016/S0370-2693(01)01060-7}{\emph{Phys. Lett.}
  {\bf B518} (2001) 8--14}, [\href{http://arxiv.org/abs/astro-ph/0012504}{{\tt
  astro-ph/0012504}}].

\bibitem{Mangano:2006mp}
G.~Mangano, A.~Melchiorri, P.~Serra, A.~Cooray and M.~Kamionkowski,
  \emph{{Cosmological bounds on dark matter-neutrino interactions}},
  \href{http://dx.doi.org/10.1103/PhysRevD.74.043517}{\emph{Phys. Rev.} {\bf
  D74} (2006) 043517}, [\href{http://arxiv.org/abs/astro-ph/0606190}{{\tt
  astro-ph/0606190}}].

\bibitem{Wilkinson:2014ksa}
R.~J. Wilkinson, C.~Boehm and J.~Lesgourgues, \emph{{Constraining Dark
  Matter-Neutrino Interactions using the CMB and Large-Scale Structure}},
  \href{http://dx.doi.org/10.1088/1475-7516/2014/05/011}{\emph{JCAP} {\bf 1405}
  (2014) 011}, [\href{http://arxiv.org/abs/1401.7597}{{\tt 1401.7597}}].

\bibitem{Frankiewicz:2015zma}
{\scshape Super-Kamiokande} collaboration, K.~Frankiewicz, \emph{{Searching for
  Dark Matter Annihilation into Neutrinos with Super-Kamiokande}},  in
  \emph{{Proceedings, Meeting of the APS Division of Particles and Fields (DPF
  2015): Ann Arbor, Michigan, USA, 4-8 Aug 2015}}, 2015.
\newblock \href{http://arxiv.org/abs/1510.07999}{{\tt 1510.07999}}.

\bibitem{Frankiewicz:2017trk}
{\scshape Super-Kamiokande} collaboration, K.~Frankiewicz, \emph{{Dark matter
  searches with the Super-Kamiokande detector}},
  \href{http://dx.doi.org/10.1088/1742-6596/888/1/012210}{\emph{J. Phys. Conf.
  Ser.} {\bf 888} (2017) 012210}.

\bibitem{Aartsen:2015xej}
{\scshape IceCube} collaboration, M.~G. Aartsen et~al., \emph{{Search for Dark
  Matter Annihilation in the Galactic Center with IceCube-79}},
  \href{http://dx.doi.org/10.1140/epjc/s10052-015-3713-1}{\emph{Eur. Phys. J.}
  {\bf C75} (2015) 492}, [\href{http://arxiv.org/abs/1505.07259}{{\tt
  1505.07259}}].

\bibitem{Albert:2016emp}
A.~Albert et~al., \emph{{Results from the search for dark matter in the Milky
  Way with 9 years of data of the ANTARES neutrino telescope}},
  \href{http://dx.doi.org/10.1016/j.physletb.2019.05.022,
  10.1016/j.physletb.2017.03.063}{\emph{Phys. Lett.} {\bf B769} (2017)
  249--254}, [\href{http://arxiv.org/abs/1612.04595}{{\tt 1612.04595}}].

\bibitem{Abdallah:2016ygi}
{\scshape H.E.S.S.} collaboration, H.~Abdallah et~al., \emph{{Search for dark
  matter annihilations towards the inner Galactic halo from 10 years of
  observations with H.E.S.S}},
  \href{http://dx.doi.org/10.1103/PhysRevLett.117.111301}{\emph{Phys. Rev.
  Lett.} {\bf 117} (2016) 111301}, [\href{http://arxiv.org/abs/1607.08142}{{\tt
  1607.08142}}].

\bibitem{Ackermann:2015zua}
{\scshape Fermi-LAT} collaboration, M.~Ackermann et~al., \emph{{Searching for
  Dark Matter Annihilation from Milky Way Dwarf Spheroidal Galaxies with Six
  Years of Fermi Large Area Telescope Data}},
  \href{http://dx.doi.org/10.1103/PhysRevLett.115.231301}{\emph{Phys. Rev.
  Lett.} {\bf 115} (2015) 231301}, [\href{http://arxiv.org/abs/1503.02641}{{\tt
  1503.02641}}].

\bibitem{Bellini:2010gn}
{\scshape Borexino} collaboration, G.~Bellini et~al., \emph{{Study of solar and
  other unknown anti-neutrino fluxes with Borexino at LNGS}},
  \href{http://dx.doi.org/10.1016/j.physletb.2010.12.030}{\emph{Phys. Lett. B}
  {\bf 696} (2011) 191--196}, [\href{http://arxiv.org/abs/1010.0029}{{\tt
  1010.0029}}].

\bibitem{Arguelles:2019ouk}
C.~A. Argüelles, A.~Diaz, A.~Kheirandish, A.~Olivares-Del-Campo, I.~Safa and
  A.~C. Vincent, \emph{{Dark Matter Annihilation to Neutrinos: An Updated,
  Consistent \& Compelling Compendium of Constraints}},
  \href{http://arxiv.org/abs/1912.09486}{{\tt 1912.09486}}.

\bibitem{Collaboration:2011jza}
{\scshape KamLAND} collaboration, A.~Gando et~al., \emph{{A study of
  extraterrestrial antineutrino sources with the KamLAND detector}},
  \href{http://dx.doi.org/10.1088/0004-637X/745/2/193}{\emph{Astrophys. J.}
  {\bf 745} (2012) 193}, [\href{http://arxiv.org/abs/1105.3516}{{\tt
  1105.3516}}].

\bibitem{PalomaresRuiz:2007eu}
S.~Palomares-Ruiz and S.~Pascoli, \emph{{Testing MeV dark matter with neutrino
  detectors}}, \href{http://dx.doi.org/10.1103/PhysRevD.77.025025}{\emph{Phys.
  Rev.} {\bf D77} (2008) 025025}, [\href{http://arxiv.org/abs/0710.5420}{{\tt
  0710.5420}}].

\bibitem{Klop:2018ltd}
N.~Klop and S.~Ando, \emph{{Constraints on MeV dark matter using neutrino
  detectors and their implication for the 21-cm results}},
  \href{http://dx.doi.org/10.1103/PhysRevD.98.103004}{\emph{Phys. Rev.} {\bf
  D98} (2018) 103004}, [\href{http://arxiv.org/abs/1809.00671}{{\tt
  1809.00671}}].

\bibitem{Rott:2011fh}
C.~Rott, T.~Tanaka and Y.~Itow, \emph{{Enhanced Sensitivity to Dark Matter
  Self-annihilations in the Sun using Neutrino Spectral Information}},
  \href{http://dx.doi.org/10.1088/1475-7516/2011/09/029}{\emph{JCAP} {\bf 1109}
  (2011) 029}, [\href{http://arxiv.org/abs/1107.3182}{{\tt 1107.3182}}].

\bibitem{Kappl:2011kz}
R.~Kappl and M.~W. Winkler, \emph{{New Limits on Dark Matter from
  Super-Kamiokande}},
  \href{http://dx.doi.org/10.1016/j.nuclphysb.2011.05.006}{\emph{Nucl. Phys.}
  {\bf B850} (2011) 505--521}, [\href{http://arxiv.org/abs/1104.0679}{{\tt
  1104.0679}}].

\bibitem{Primulando:2017kxf}
R.~Primulando and P.~Uttayarat, \emph{{Dark Matter-Neutrino Interaction in
  Light of Collider and Neutrino Telescope Data}},
  \href{http://dx.doi.org/10.1007/JHEP06(2018)026}{\emph{JHEP} {\bf 06} (2018)
  026}, [\href{http://arxiv.org/abs/1710.08567}{{\tt 1710.08567}}].

\bibitem{Arguelles:2019xgp}
C.~A. Argüelles et~al., \emph{{White Paper on New Opportunities at the
  Next-Generation Neutrino Experiments (Part 1: BSM Neutrino Physics and Dark
  Matter)}},  \href{http://arxiv.org/abs/1907.08311}{{\tt 1907.08311}}.

\bibitem{Abe:2018uyc}
{\scshape Hyper-Kamiokande} collaboration, K.~Abe et~al.,
  \emph{{Hyper-Kamiokande Design Report}},
  \href{http://arxiv.org/abs/1805.04163}{{\tt 1805.04163}}.

\bibitem{Beacom:2010kk}
J.~F. Beacom, \emph{{The Diffuse Supernova Neutrino Background}},
  \href{http://dx.doi.org/10.1146/annurev.nucl.010909.083331}{\emph{Ann. Rev.
  Nucl. Part. Sci.} {\bf 60} (2010) 439--462},
  [\href{http://arxiv.org/abs/1004.3311}{{\tt 1004.3311}}].

\bibitem{Mirizzi:2015eza}
A.~Mirizzi, I.~Tamborra, H.-T. Janka, N.~Saviano, K.~Scholberg, R.~Bollig
  et~al., \emph{{Supernova Neutrinos: Production, Oscillations and Detection}},
  \href{http://dx.doi.org/10.1393/ncr/i2016-10120-8}{\emph{Riv. Nuovo Cim.}
  {\bf 39} (2016) 1--112}, [\href{http://arxiv.org/abs/1508.00785}{{\tt
  1508.00785}}].

\bibitem{Bays:2011si}
{\scshape Super-Kamiokande} collaboration, K.~Bays et~al., \emph{{Supernova
  Relic Neutrino Search at Super-Kamiokande}},
  \href{http://dx.doi.org/10.1103/PhysRevD.85.052007}{\emph{Phys. Rev.} {\bf
  D85} (2012) 052007}, [\href{http://arxiv.org/abs/1111.5031}{{\tt
  1111.5031}}].

\bibitem{Zhang:2013tua}
{\scshape Super-Kamiokande} collaboration, H.~Zhang et~al., \emph{{Supernova
  Relic Neutrino Search with Neutron Tagging at Super-Kamiokande-IV}},
  \href{http://dx.doi.org/10.1016/j.astropartphys.2014.05.004}{\emph{Astropart.
  Phys.} {\bf 60} (2015) 41--46}, [\href{http://arxiv.org/abs/1311.3738}{{\tt
  1311.3738}}].

\bibitem{Beacom:2003nk}
J.~F. Beacom and M.~R. Vagins, \emph{{GADZOOKS! Anti-neutrino spectroscopy with
  large water Cherenkov detectors}},
  \href{http://dx.doi.org/10.1103/PhysRevLett.93.171101}{\emph{Phys. Rev.
  Lett.} {\bf 93} (2004) 171101},
  [\href{http://arxiv.org/abs/hep-ph/0309300}{{\tt hep-ph/0309300}}].

\bibitem{Horiuchi:2008jz}
S.~Horiuchi, J.~F. Beacom and E.~Dwek, \emph{{The Diffuse Supernova Neutrino
  Background is detectable in Super-Kamiokande}},
  \href{http://dx.doi.org/10.1103/PhysRevD.79.083013}{\emph{Phys. Rev.} {\bf
  D79} (2009) 083013}, [\href{http://arxiv.org/abs/0812.3157}{{\tt
  0812.3157}}].

\bibitem{Campo:2018dfh}
A.~Olivares-Del~Campo, S.~Palomares-Ruiz and S.~Pascoli, \emph{{Implications of
  a Dark Matter-Neutrino Coupling at Hyper-Kamiokande}},  in
  \emph{{Proceedings, 53rd Rencontres de Moriond on Electroweak Interactions
  and Unified Theories (Moriond EW 2018): La Thuile, Italy, March 10-17,
  2018}}, pp.~441--444, 2018.
\newblock \href{http://arxiv.org/abs/1805.09830}{{\tt 1805.09830}}.

\bibitem{Navarro:1995iw}
J.~F. Navarro, C.~S. Frenk and S.~D.~M. White, \emph{{The Structure of cold
  dark matter halos}}, \href{http://dx.doi.org/10.1086/177173}{\emph{Astrophys.
  J.} {\bf 462} (1996) 563--575},
  [\href{http://arxiv.org/abs/astro-ph/9508025}{{\tt astro-ph/9508025}}].

\bibitem{Moore:1999gc}
B.~Moore, T.~R. Quinn, F.~Governato, J.~Stadel and G.~Lake, \emph{{Cold
  collapse and the core catastrophe}},
  \href{http://dx.doi.org/10.1046/j.1365-8711.1999.03039.x}{\emph{Mon. Not.
  Roy. Astron. Soc.} {\bf 310} (1999) 1147--1152},
  [\href{http://arxiv.org/abs/astro-ph/9903164}{{\tt astro-ph/9903164}}].

\bibitem{Bahcall:1980}
J.~N. {Bahcall} and R.~M. {Soneira}, \emph{{The universe at faint magnitudes.
  I. Models for the Galaxy and the predicted star counts.}},
  \href{http://dx.doi.org/10.1086/190685}{\emph{ApJS} {\bf 44} (Sep, 1980)
  73--110}.

\bibitem{Jiang:2019xwn}
{\scshape Super-Kamiokande} collaboration, M.~Jiang et~al., \emph{{Atmospheric
  Neutrino Oscillation Analysis with Improved Event Reconstruction in
  Super-Kamiokande IV}},
  \href{http://dx.doi.org/10.1093/ptep/ptz015}{\emph{PTEP} {\bf 2019} (2019)
  053F01}, [\href{http://arxiv.org/abs/1901.03230}{{\tt 1901.03230}}].

\bibitem{Brun:1997pa}
R.~Brun and F.~Rademakers, \emph{{ROOT: An object oriented data analysis
  framework}},
  \href{http://dx.doi.org/10.1016/S0168-9002(97)00048-X}{\emph{Nucl. Instrum.
  Meth.} {\bf A389} (1997) 81--86}.

\bibitem{Fukuda:2002uc}
{\scshape Super-Kamiokande} collaboration, Y.~Fukuda et~al., \emph{{The
  Super-Kamiokande detector}},
  \href{http://dx.doi.org/10.1016/S0168-9002(03)00425-X}{\emph{Nucl. Instrum.
  Meth.} {\bf A501} (2003) 418--462}.

\bibitem{Abe:2013gga}
K.~Abe et~al., \emph{{Calibration of the Super-Kamiokande Detector}},
  \href{http://dx.doi.org/10.1016/j.nima.2013.11.081}{\emph{Nucl. Instrum.
  Meth.} {\bf A737} (2014) 253--272},
  [\href{http://arxiv.org/abs/1307.0162}{{\tt 1307.0162}}].

\bibitem{Brun:1994aa}
R.~Brun, F.~Bruyant, F.~Carminati, S.~Giani, M.~Maire, A.~McPherson et~al.,
  \emph{{GEANT Detector Description and Simulation Tool}}, .

\bibitem{Andreopoulos:2009rq}
C.~Andreopoulos et~al., \emph{{The GENIE Neutrino Monte Carlo Generator}},
  \href{http://dx.doi.org/10.1016/j.nima.2009.12.009}{\emph{Nucl. Instrum.
  Meth.} {\bf A614} (2010) 87--104},
  [\href{http://arxiv.org/abs/0905.2517}{{\tt 0905.2517}}].

\bibitem{Andreopoulos:2015wxa}
C.~Andreopoulos, C.~Barry, S.~Dytman, H.~Gallagher, T.~Golan, R.~Hatcher
  et~al., \emph{{The GENIE Neutrino Monte Carlo Generator: Physics and User
  Manual}},  \href{http://arxiv.org/abs/1510.05494}{{\tt 1510.05494}}.

\bibitem{Honda:2011nf}
M.~Honda, T.~Kajita, K.~Kasahara and S.~Midorikawa, \emph{{Improvement of low
  energy atmospheric neutrino flux calculation using the JAM nuclear
  interaction model}},
  \href{http://dx.doi.org/10.1103/PhysRevD.83.123001}{\emph{Phys. Rev.} {\bf
  D83} (2011) 123001}, [\href{http://arxiv.org/abs/1102.2688}{{\tt
  1102.2688}}].

\bibitem{Battistoni:2005pd}
G.~Battistoni, A.~Ferrari, T.~Montaruli and P.~R. Sala, \emph{{The atmospheric
  neutrino flux below 100-MeV: The FLUKA results}},
  \href{http://dx.doi.org/10.1016/j.astropartphys.2005.03.006}{\emph{Astropart.
  Phys.} {\bf 23} (2005) 526--534}.

\bibitem{Gondolo:2004sc}
P.~Gondolo, J.~Edsjo, P.~Ullio, L.~Bergstrom, M.~Schelke and E.~A. Baltz,
  \emph{{DarkSUSY: Computing supersymmetric dark matter properties
  numerically}},
  \href{http://dx.doi.org/10.1088/1475-7516/2004/07/008}{\emph{JCAP} {\bf 0407}
  (2004) 008}, [\href{http://arxiv.org/abs/astro-ph/0406204}{{\tt
  astro-ph/0406204}}].

\bibitem{Bringmann:2018lay}
T.~Bringmann, J.~Edsjö, P.~Gondolo, P.~Ullio and L.~Bergström,
  \emph{{DarkSUSY 6 : An Advanced Tool to Compute Dark Matter Properties
  Numerically}},
  \href{http://dx.doi.org/10.1088/1475-7516/2018/07/033}{\emph{JCAP} {\bf 1807}
  (2018) 033}, [\href{http://arxiv.org/abs/1802.03399}{{\tt 1802.03399}}].

\bibitem{Wallraff:2014qka}
M.~Wallraff and C.~Wiebusch, \emph{{Calculation of oscillation probabilities of
  atmospheric neutrinos using nuCraft}},
  \href{http://dx.doi.org/10.1016/j.cpc.2015.07.010}{\emph{Comput. Phys.
  Commun.} {\bf 197} (2015) 185--189},
  [\href{http://arxiv.org/abs/1409.1387}{{\tt 1409.1387}}].

\bibitem{Dziewonski:1981xy}
A.~M. Dziewonski and D.~L. Anderson, \emph{{Preliminary reference earth
  model}}, \href{http://dx.doi.org/10.1016/0031-9201(81)90046-7}{\emph{Phys.
  Earth Planet. Interiors} {\bf 25} (1981) 297--356}.

\bibitem{Tanabashi:2018oca}
{\scshape Particle Data Group} collaboration, M.~Tanabashi et~al.,
  \emph{{Review of Particle Physics}},
  \href{http://dx.doi.org/10.1103/PhysRevD.98.030001}{\emph{Phys. Rev.} {\bf
  D98} (2018) 030001}.

\bibitem{Martini:2009uj}
M.~Martini, M.~Ericson, G.~Chanfray and J.~Marteau, \emph{{A Unified approach
  for nucleon knock-out, coherent and incoherent pion production in neutrino
  interactions with nuclei}},
  \href{http://dx.doi.org/10.1103/PhysRevC.80.065501}{\emph{Phys. Rev.} {\bf
  C80} (2009) 065501}, [\href{http://arxiv.org/abs/0910.2622}{{\tt
  0910.2622}}].

\bibitem{LlewellynSmith:1971uhs}
C.~H. Llewellyn~Smith, \emph{{Neutrino Reactions at Accelerator Energies}},
  \href{http://dx.doi.org/10.1016/0370-1573(72)90010-5}{\emph{Phys. Rept.} {\bf
  3} (1972) 261--379}.

\bibitem{Nieves:2004wx}
J.~Nieves, J.~E. Amaro and M.~Valverde, \emph{{Inclusive quasi-elastic neutrino
  reactions}}, \href{http://dx.doi.org/10.1103/PhysRevC.70.055503,
  10.1103/PhysRevC.72.019902}{\emph{Phys. Rev.} {\bf C70} (2004) 055503},
  [\href{http://arxiv.org/abs/nucl-th/0408005}{{\tt nucl-th/0408005}}].

\bibitem{Nieves:2005}
J.~{Nieves}, J.~E. {Amaro} and M.~{Valverde}, \emph{{Erratum: Inclusive
  quasielastic charged-current neutrino-nucleus reactions [Phys. Rev. C 70,
  055503 (2004)]}},
  \href{http://dx.doi.org/10.1103/PhysRevC.72.019902}{\emph{Phys. Rev.} {\bf
  C72} (2005) 019902}.

\bibitem{Nieves:2011pp}
J.~Nieves, I.~Ruiz~Simo and M.~J. Vicente~Vacas, \emph{{Inclusive
  Charged--Current Neutrino--Nucleus Reactions}},
  \href{http://dx.doi.org/10.1103/PhysRevC.83.045501}{\emph{Phys. Rev.} {\bf
  C83} (2011) 045501}, [\href{http://arxiv.org/abs/1102.2777}{{\tt
  1102.2777}}].

\bibitem{Li:2014sea}
S.~W. Li and J.~F. Beacom, \emph{{First calculation of cosmic-ray muon
  spallation backgrounds for MeV astrophysical neutrino signals in
  Super-Kamiokande}},
  \href{http://dx.doi.org/10.1103/PhysRevC.89.045801}{\emph{Phys. Rev.} {\bf
  C89} (2014) 045801}, [\href{http://arxiv.org/abs/1402.4687}{{\tt
  1402.4687}}].

\bibitem{Li:2015kpa}
S.~W. Li and J.~F. Beacom, \emph{{Spallation Backgrounds in Super-Kamiokande
  Are Made in Muon-Induced Showers}},
  \href{http://dx.doi.org/10.1103/PhysRevD.91.105005}{\emph{Phys. Rev.} {\bf
  D91} (2015) 105005}, [\href{http://arxiv.org/abs/1503.04823}{{\tt
  1503.04823}}].

\bibitem{Li:2015lxa}
S.~W. Li and J.~F. Beacom, \emph{{Tagging Spallation Backgrounds with Showers
  in Water-Cherenkov Detectors}},
  \href{http://dx.doi.org/10.1103/PhysRevD.92.105033}{\emph{Phys. Rev.} {\bf
  D92} (2015) 105033}, [\href{http://arxiv.org/abs/1508.05389}{{\tt
  1508.05389}}].

\bibitem{Super-Kamiokande:2015xra}
{\scshape Super-Kamiokande} collaboration, Y.~Zhang et~al., \emph{{First
  measurement of radioactive isotope production through cosmic-ray muon
  spallation in Super-Kamiokande IV}},
  \href{http://dx.doi.org/10.1103/PhysRevD.93.012004}{\emph{Phys. Rev.} {\bf
  D93} (2016) 012004}, [\href{http://arxiv.org/abs/1509.08168}{{\tt
  1509.08168}}].

\bibitem{Groom:2001kq}
D.~E. Groom, N.~V. Mokhov and S.~I. Striganov, \emph{{Muon stopping power and
  range tables 10-MeV to 100-TeV}},
  \href{http://dx.doi.org/10.1006/adnd.2001.0861}{\emph{Atom. Data Nucl. Data
  Tabl.} {\bf 78} (2001) 183--356}.

\bibitem{Tsai:1973py}
Y.-S. Tsai, \emph{{Pair Production and Bremsstrahlung of Charged Leptons}},
  \href{http://dx.doi.org/10.1103/RevModPhys.46.815,
  10.1103/RevModPhys.49.421}{\emph{Rev. Mod. Phys.} {\bf 46} (1974) 815}.

\bibitem{wcsim}
T.~Dealtry, A.~Himmel, J.~Hoppenau and J.~Lozier, ``{Water Cherenkov Simulator
  (WCSim)}.'' \url{https://github.com/WCSim/WCSim}, 2016.

\bibitem{Ashie:2005ik}
{\scshape Super-Kamiokande} collaboration, Y.~Ashie et~al., \emph{{A
  Measurement of atmospheric neutrino oscillation parameters by
  SUPER-KAMIOKANDE I}},
  \href{http://dx.doi.org/10.1103/PhysRevD.71.112005}{\emph{Phys. Rev.} {\bf
  D71} (2005) 112005}, [\href{http://arxiv.org/abs/hep-ex/0501064}{{\tt
  hep-ex/0501064}}].

\bibitem{Hayato:2002sd}
Y.~Hayato, \emph{{NEUT}},
  \href{http://dx.doi.org/10.1016/S0920-5632(02)01759-0}{\emph{Nucl. Phys. B
  Proc. Suppl.} {\bf 112} (2002) 171--176}.

\bibitem{Abe:2017aap}
{\scshape Super-Kamiokande} collaboration, K.~Abe et~al., \emph{{Atmospheric
  neutrino oscillation analysis with external constraints in Super-Kamiokande
  I-IV}}, \href{http://dx.doi.org/10.1103/PhysRevD.97.072001}{\emph{Phys. Rev.}
  {\bf D97} (2018) 072001}, [\href{http://arxiv.org/abs/1710.09126}{{\tt
  1710.09126}}].

\bibitem{Gordon:2013vta}
C.~Gordon and O.~Macias, \emph{{Dark Matter and Pulsar Model Constraints from
  Galactic Center Fermi-LAT Gamma Ray Observations}},
  \href{http://dx.doi.org/10.1103/PhysRevD.88.083521,
  10.1103/PhysRevD.89.049901}{\emph{Phys. Rev.} {\bf D88} (2013) 083521},
  [\href{http://arxiv.org/abs/1306.5725}{{\tt 1306.5725}}].

\bibitem{Edwards:2017mnf}
T.~D.~P. Edwards and C.~Weniger, \emph{{A Fresh Approach to Forecasting in
  Astroparticle Physics and Dark Matter Searches}},
  \href{http://dx.doi.org/10.1088/1475-7516/2018/02/021}{\emph{JCAP} {\bf 1802}
  (2018) 021}, [\href{http://arxiv.org/abs/1704.05458}{{\tt 1704.05458}}].

\bibitem{Edwards:2017kqw}
T.~D.~P. Edwards and C.~Weniger, \emph{{swordfish: Efficient Forecasting of New
  Physics Searches without Monte Carlo}},
  \href{http://arxiv.org/abs/1712.05401}{{\tt 1712.05401}}.

\bibitem{Steigman:2012nb}
G.~Steigman, B.~Dasgupta and J.~F. Beacom, \emph{{Precise Relic WIMP Abundance
  and its Impact on Searches for Dark Matter Annihilation}},
  \href{http://dx.doi.org/10.1103/PhysRevD.86.023506}{\emph{Phys. Rev.} {\bf
  D86} (2012) 023506}, [\href{http://arxiv.org/abs/1204.3622}{{\tt
  1204.3622}}].

\bibitem{Abe:2016ero}
{\scshape Hyper-Kamiokande} collaboration, K.~Abe et~al., \emph{{Physics
  potentials with the second Hyper-Kamiokande detector in Korea}},
  \href{http://dx.doi.org/10.1093/ptep/pty044}{\emph{PTEP} {\bf 2018} (2018)
  063C01}, [\href{http://arxiv.org/abs/1611.06118}{{\tt 1611.06118}}].

\bibitem{Reid:2004rd}
M.~J. Reid and A.~Brunthaler, \emph{{The Proper motion of Sgr A*. 2. The Mass
  of Sgr A*}}, \href{http://dx.doi.org/10.1086/424960}{\emph{Astrophys. J.}
  {\bf 616} (2004) 872--884},
  [\href{http://arxiv.org/abs/astro-ph/0408107}{{\tt astro-ph/0408107}}].

\bibitem{astropy:2013}
{Astropy Collaboration}, T.~P. {Robitaille}, E.~J. {Tollerud}, P.~{Greenfield},
  M.~{Droettboom}, E.~{Bray} et~al., \emph{{Astropy: A community Python package
  for astronomy}},  \href{http://arxiv.org/abs/1307.6212}{{\tt 1307.6212}}.

\bibitem{astropy:2018}
A.~M. {Price-Whelan}, B.~M. {Sip{\H{o}}cz}, H.~M. {G{\"u}nther}, P.~L. {Lim},
  S.~M. {Crawford}, S.~{Conseil} et~al., \emph{{The Astropy Project: Building
  an Open-science Project and Status of the v2.0 Core Package}}, .

\bibitem{SOFA:2019-07-22}
{IAU SOFA Board},   \emph{{IAU SOFA Software Collection}}, \href{http://www.iausofa.org}{http://www.iausofa.org}.

\end{thebibliography}\endgroup

\end{document}